\documentclass[authoryear,12pt]{elsarticle}
\usepackage{lineno,hyperref}
\modulolinenumbers[5]

\usepackage[paperwidth=8.5 in,paperheight=11 in,twoside,textwidth=6.5 in,left=1 in, right = 1 in]{geometry} 

\makeatletter
\def\ps@pprintTitle{%
	\let\@oddhead\@empty
	\let\@evenhead\@empty
	\def\@oddfoot{\centerline{\it{}}}%
	\let\@evenfoot\@oddfoot}
\makeatother

\usepackage{natbib}
\usepackage{subfig} 
\usepackage{epstopdf} 
\usepackage{amsmath,amssymb,amsthm,amsfonts} 
\usepackage{varioref}
\usepackage{doi}
\usepackage{booktabs}
\usepackage{caption}
\usepackage{multirow}
\usepackage{longtable}
\usepackage{pdflscape}
\usepackage{xcolor}
\usepackage{lscape}

\usepackage{setspace} 

\usepackage{verbatim}

\usepackage{adjustbox}
\usepackage{float}

\usepackage{booktabs}

\theoremstyle{definition}

\theoremstyle{plain}

\theoremstyle{plain}

\theoremstyle{plain}

\theoremstyle{plain}

\begin{document}
	\onehalfspacing
	\begin{frontmatter}
		
	\title{Cyber Risk Taxonomies: Statistical Analysis of Cybersecurity Risk Classifications}

  \author[add4]{Matteo Malavasi\corref{cor1}}
  \ead{m.malavasi@unsw.edu.au}  
  \author[add2]{Gareth W. Peters}
  \ead{garethpeters@ucsb.edu} 
  \author[add1]{Stefan Tr\"uck}
  \ead{Stefan.Trueck@mq.edu.au}  
  \author[add1]{Pavel V. Shevchenko}
  \ead{Pavel.Shevchenko@mq.edu.au}  
  \author[add1]{Jiwook Jang}
  \ead{Jiwook.Jang@mq.edu.au}
  \author[add3]{Georgy Sofronov}
  \ead{Georgy.Sofronov@mq.edu.au}

  \cortext[cor1]{Please address correspondence to Matteo Malavasi}
  \address[add4]{School of Risk and Actuarial Studies, UNSW Business School, University of New South Wales, Australia}
  \address[add1]{Department of Actuarial Studies and Business Analytics, Macquarie University, Australia}
  \address[add2]{Department of Statistics and Applied Probability, University of California Santa Barbara, USA}
  \address[add3]{School of Mathematical and Physical Sciences, Macquarie University, Australia}

\begin{abstract}
Cyber risk classifications are widely used in the modeling of cyber event distributions, yet their effectiveness in out of sample forecasting performance remains underexplored. In this paper, we analyse the most commonly used classifications and argue in favour of switching the attention from goodness-of-fit and in-sample predictive performance, to focusing on the out-of sample forecasting performance. We use a rolling window analysis, to compare cyber risk distribution forecasts via threshold weighted scoring functions. Our results indicate that business motivated cyber risk classifications appear to be too restrictive and not flexible enough to capture the heterogeneity of cyber risk events. We investigate how dynamic and impact-based cyber risk classifiers seem to be better suited in forecasting future cyber risk losses than the other considered classifications. These findings suggest that cyber risk types provide limited forecasting ability concerning cyber event severity distribution, and cyber insurance ratemakers should utilize cyber risk types only when modeling the cyber event frequency distribution. Our study offers valuable insights for decision-makers and policymakers alike, contributing to the advancement of scientific knowledge in the field of cyber risk management.
\end{abstract}

\end{frontmatter}
	

\section{Introduction}

In an era where cyber threats continue to evolve and proliferate, accurate assessment and management of cyber risks have become a paramount concern for organizations across various sectors. Cyber risk classifications play a crucial role in the identification, measurement, modeling, and management of cyber risk events, providing structured frameworks to understand and categorize different types of cyber threats \citep[see, e.g.][]{pub2004standards, nai2018,nai2019, acsc2020, eling2019,romanosky2016, curti2023cyber}. However, classifying cyber risk presents a formidable challenge due to its diverse array of risk factors, its dynamic and complex nature, and its pervasive impact in both the public and private sectors \citep{peters2018}. Cyber risk affects a multifaceted landscape, with diverse technical expertise, behavioral tendencies, and cultural aspects. Individuals, private enterprises, and public entities face a myriad of consequences, ranging from data breaches to ransomware attacks, credit card fraud, and identity theft.  The nature of these attacks is highly variable and poses challenges in both detection and prevention over time and across different targets. The heterogeneous nature of cyber risk defies a unique and globally accepted classification framework, leading to the development of diverse taxonomies tailored to specific industry perspectives \cite[see, e.g.][]{cebula2010,cebula2014,pub2004standards, romanosky2016, cro2017cro,cyentia2020}. Consequently, numerous public and private institutions have formulated classifications designed for the distinct needs of their stakeholders \citep{pub2004standards,nai2018, acsc2020}. Regulatory bodies have also issued cyber risk classifications aimed at various facets of cyber risk management.

In this study, we analyse several commonly employed approaches for classifying cyber risks by focusing on the forecasting abilities of these classification methodologies. Specifically, we argue in favor of prioritizing the out of sample forecasting performance as a metric for evaluating cyber risk classification models, as opposed to solely relying on in-sample goodness-of-fit measures. In modeling cyber risk event for forecasting purposes there are two main challenges. Firstly, one needs to face the scarcity of data that permeates the field of research on cyber risk \cite[see, e.g.][]{maillart2010,eling2019,aldasoro2020, malavasi2022,  curti2023cyber}. There are only a handful of datasets that could be considered suitable for research purposes, and that are available to the public\footnote{Available datasets can be obtained by different providers. Vendors include, for example, Privacy Clearing House (\url{https://privacyrights.org/data-breaches}), ORX (\url{https://orx.org/cyber}), CyberDB (\url{https://cyberdb.co/}), and Verisk (\url{https://verisk.com/}).}. We investigate forecasting ability of cyber risk classification by leveraging on the Advisen cyber loss dataset (\url{https://advisenltd.com/}), which contains information on cyber event related losses,  suffered by companies in the United States and across the globe, reported in monetary value \cite[see,][]{aldasoro2020,malavasi2022, shevchenko2023, peters2023}. One of the advantages of basing our analysis on the Advisen cyber loss data, is that it contains measurement of direct losses caused by cyber risk events, measured in dollars, therefore providing the information needed in estimating the cyber event severity distribution. Second, it is well documented in the literature that the distribution of cyber events suffers from the presence of extreme events, rendering regression framework based on Ordinary Least Squares (OLS) inadequate to accurately capture the severity of cyber events 
\citep{maillart2010,eling2017,malavasi2022,peters2023,shevchenko2023}. To address these challenges, we adopt an approach based on dynamic extreme value theory (EVT) combined with the Generalized Additive Model for Location, Scale and Shape (GAMLSS) regression framework \cite[see, e.g.][]{rigby2005,stasinopoulos2008,chavez2016,stasinopoulos2017}. In particular, to model cyber event severity distributions, we adopt the peaks-over-threshold method (POT), where exceedances over a high enough threshold can be assumed to follow a generalized Pareto distribution. This method has proven to be very efficient in capturing the risk profile of the distribution of cyber events by focusing on the tail of the distribution \cite[see, e.g.][]{eling2019,malavasi2022}. To capture the intrinsic heterogeneous nature of cyber risk, we allow the distributional parameters to depend on covariates via the GAMLSS framework.
 
 In order to evaluate the forecasting ability of a specific cyber risk classification from a distributional point of view, we adopt the paradigm of maximizing the sharpness of the predictive distributions subject to calibration as well as additional scoring rules \cite[see,][]{gneiting2007A,gneiting2007B,gneiting2011}. Specifically, we use threshold-weighted scoring rules, including the Continuous Ranked Probability Score (CRPS) and the Energy Score (ES), to rigorously test hypotheses regarding the forecasting performance of modeling the severity distribution for cyber events based on different classifications of cyber risk. To account for the presence of heavy tail behavior and an infinite mean loss process that is typical for cyber event related loss profiles, we modify the testing methodology in \cite{amisano2007} and \cite{gneiting2011} and introduce the residual Continuously Ranked Probability Score (rCRPS) and the residual Energy Score (rES). The rCRPS and rES combine scoring rules, with the standardized residuals typical of the GAMLSS approach and allow the testing procedure to be carried out even in the case of a cyber event severity distribution without finite moments. We also proceed to investigate whether cyber risk classifications have any discriminating power concerning the frequency distribution of cyber events. Finally, we propose a simulation study to investigate how robust our results are with respect to the sample size and the statistical power of our testing methodology.

 Our analysis offers multiple insights on the performance of cyber risk classifications and their impact on modeling the severity distribution of cyber events. First, we find that cyber event severity models, estimated based on cyber risk classifications have an overall unsatisfactory forecasting performance, often not statistically distinguishable from a model with a random classifications or from a model without any classification. This may suggest that empirical results based on statistically significant variables, reflecting these risk types, should be interpreted and used with care in forecasting cyber event severity \cite[see, e.g.][]{romanosky2016,aldasoro2022}. Nonetheless, the usage of cyber risk types in cyber risk management and other business motivated purposes remains a valid approach \citep{shetty2018,eling2021,gatzert2022}. Second, while cyber risk classifications may be disregarded in severity modeling, they appear to be a valuable tool for modeling the frequency of cyber events. This has important implications for cyber insurance ratemakers, since it suggests to use a classification-free model for severity, and then to adjust premium calculations according to the frequency distribution pending on the adopted risk classification.
 Moreover, our results suggest that considering cyber risk as a subcategory of operational risk provides useful insights only limited to mandatory disclosures and best cyber risk management practices, but it fails to return a satisfactory out of sample forecasting performance. These results are in line with previous findings in the literature, pointing out the difference between operational risk and cyber risk events from a statistical point of view \citep{eling2019, malavasi2022}. Finally, implementing an additional simulation study, we show how the testing procedure of the forecasting performance in the context of cyber risk may suffer from low statistical power, but the inclusion of a weighting regime in the scoring rules can help with increasing the level of statistical power, even in small samples.
 Our paper relates to the stream of literature on statistical cyber risk modeling \citep{eling2015modelling,eling2017,eling2018,eling2019, eling2020cyber,jung2021,malavasi2022,peters2023,shevchenko2023}, and the one on cyber risk classifications \citep{pub2004standards,romanosky2016,healey2018future,nai2018,peters2018,shetty2018,duffie2019cyber,nai2019,romanosky2019content,acsc2020,eling2021,eisenbach2022cyber,gatzert2022,curti2023cyber}. We also contribute to the discussion on whether cyber risk should be considered as a subcategory of operational risk, see, e.g. \cite{cebula2010,cebula2014,biener2015,eling2018,cohen2019,eling2019,kashyap2019some,adelmann2020cyber, aldasoro2022,gatzert2022}.

The remainder of the manuscript is organized as follows. Section \ref{sec:model} introduces the model set up that is applied in the empirical analysis. Section \ref{sec:class} discusses the cyber risk classification approaches considered in this study. Section \ref{sec:data} describes the Advisen Cyber Loss dataset. Section \ref{sec:results} presents the results of the empirical analysis and relates them to previous findings in the literature. Finally, Section \ref{sec:con} concludes and discusses implications of our findings for the classification and management of cyber risks.

\section{Model Set Up}
\label{sec:model}
Cyber event related losses are well known to show extreme events and heterogeneity \cite[see, e.g.][]{eling2019,malavasi2022,peters2023,shevchenko2023}. We use a combined estimation approach based on the POT method from extreme value theory and the GAMLSS framework to estimate the severity distribution of cyber events. 

We then use the estimated severity distribution in a series of hypothesis tests to assess the forecasting performance of most commonly used classifications, based on the paradigm of maximising the sharpness of the predictive distribution subject to calibration as well as the application of different scoring rules \cite[see,][]{gneiting2007A,gneiting2007B,gneiting2011}. We consider both cyber event severity and frequency distributions. By focusing on the tail of the severity distribution, we can benefit from a combination of methods from extreme value theory and the GAMLSS approach  that has been proven to provide a good fit to losses from cyber events in the literature \cite[see, e.g.][]{maillart2010,eling2017,eling2019,jung2021,malavasi2022}. Second, we analyze the frequency distribution of cyber events in the latter part of the manuscript to provide useful insights for the rate-making function of insurers. 

\subsection{Cyber Event Severity}
Let the $\widetilde{Y}_{i,t}$ be the $i$-th loss during year $t$, and let $X_{i,t}$ be a vector a of company and event-specific characteristics.  We assume that $\widetilde{Y}_{i,t}$, $i=1,\dots,\widetilde{N}_t$, $t=1,\dots,T$ are independent and follow a continuous distribution $F_{\widetilde{Y}}$. Under the assumption that 
 $F_{\widetilde{Y}}$ satisfies the following regular variation condition:
\begin{equation}
\label{eq:varying}
    \Bar{F}_{\widetilde{Y}}(x) = 1-F_{\widetilde{Y}}(x) \sim x^{-\tau}L(x),\;x\rightarrow \infty, \tau>0
\end{equation}
for some measurable, slowly varying function $L:(0,\infty)\to(0,\infty)$\cite[see,][]{balkema1974,pickands1975}, the exceedances $Y_{i,t} = \widetilde{Y}_{i,t}-u|\widetilde{Y}_{i,t}>u,$ for a high enough threshold $u$, follow the generalized Pareto distribution with the following density \cite[see,][]{pickands1975}:
\begin{equation}
\label{eq:pareto_density}
    g(y;\mu,\tau)=\frac{\mu}{\tau}\left(1+\frac{y}{\mu} \right)^{-(1+\tau)}
\end{equation}
for $y\ge 0$ 
\cite[see, e.g.][]{ganegoda2013,chavez2016}. To better capture the dependence of each distributional parameter on the covariates, we adopt the GAMLSS approach and allow each distributional parameter in Equation (\ref{eq:pareto_density}) to depend on covariates via the following set of link functions \citep{rigby2005,stasinopoulos2008,ganegoda2013,eling2019,malavasi2022}:
\begin{align}
& \log(\mu(X_{i,t},t)) = f_\mu(X_{i,t}) + h_\mu(t),\notag\\
& \log(\tau(X_{i,t},t)) = f_\tau(X_{i,t})+h_\tau(t),\notag
\end{align}
where $h_\xi,h_\nu$ are measurable, twice differentiable functions, and $X_{i,t}$ is a vector of company and event specific characteristics corresponding to the event $i$ at time $t$. The functional form of $f_\mu$ and $f_\tau$ is assumed to be linear, while the choice of $h_\mu,h_\tau$ is made using information criteria. Under the assumption of independence of $Y_{i,t}$, $i =1,\dots,N_t$, $t=1,\dots,T$ the estimation can be carried out via the following penalized maximum likelihood \cite[see,][]{chavez2016}:
\begin{equation}
\label{eq:likelihood}
    \max \sum_{t=1}^{T}\sum_{i=1}^{N_t}\log\left(g \left(Y_{i,t};\mu(X_{i,t},t),\tau(X_{i,t},t)\right)\right) - \gamma_{\mu}\int_0^T h''_\mu(s)ds - \gamma_\tau\int_0^Th_\tau''(s)ds,
\end{equation}
where $\gamma_\mu$ and $\gamma_\tau$ are the smoothing parameters.

\subsection{Scoring Rules and Distributional Forecasts}
Scoring rules can be used to compare distributional forecasts with realizations via a loss function $S(F, y)$, where $F$ is the cumulative density forecast and $y$ is the future realization \cite[see, e.g. ][]{gneiting2007A,gneiting2007B,gneiting2011}. Let $\mathcal{F}$ be the convex class of distributions defined on $\left(\Omega,\mathcal{A}\right)$, $y$ be the realisation of a random variable $Y$, and $F$ be the forecast of the distribution of $Y$. A scoring rule is a function such that $S(F,y):\mathcal{F}\times\omega \to \mathbb{R}$. Given the true distribution of $Y$, a scoring rule $F^*$ is proper if:
\begin{equation}
\label{eq:proper}
    \mathbb{E}_{F^*}\left[S(F^*,.)\right] = \int S(F^*,.)dF^*\geq \int S(F,.)dF^* = \mathbb{E}_{F^*}\left[S(F,.)\right]
\end{equation}
for all $F\in\mathcal{F}$. Moreover, if $F^*$ is the unique maximizer of Equation (\ref{eq:proper}), then $S$ is also a strict scoring rule, \cite[see,][]{gneiting2007A,gneiting2007B,rizzo2009,gneiting2011,szekely2013,alexander2022}.
The most commonly used scoring rule is the threshold weighted continuously ranked probability score (twCRPS):
\begin{equation}
\label{eq:crps}
    S(F,y) = \int_{-\infty}^{\infty}PS\left(F(z), \mathbb{I}_{y\leq z} \right) u(z)dz,\notag
\end{equation}
where $F(z) = \int_{-\infty}^z f(y)dy$, $\mathbb{I}_{y\leq z}$ is the indicator function being equal to 1 is $y\leq z$, and equal to 0 otherwise,
\begin{equation}
    PS\left(F(z), \mathbb{I}_{y\leq z} \right)  = \left(F(z)- \mathbb{I}_{y\leq z} \right)^2,
\end{equation}
is the Brier probability score for the probability estimate $F(z)$ at the binary event $\{Y\leq z\}$ \cite[see, ][]{selten1998, gneiting2007B}, and $u(z)$ is a non negative weighting function on the real line. The scoring rule in Equation (\ref{eq:crps}) measures predictive performance of the probability distribution $F$ that one wishes to maximize, i.e., a density forecast with a higher $S(F,y)$ is preferred. By selecting different weighting functions $u$, the attention can be posed on a specific part of the distribution, such as the body, or the tail. When $u(z)=1$ for all $ z\in \mathbb{R}$, $S(F,y)$ is equivalent to the continuously ranked probability score (CRPS) \citep{gneiting2011}. The weighting function $u$ can be arbitrarily chosen, provided that it's positive and bounded. Following the suggestion in \cite{amisano2007} and \cite{gneiting2011} we use the following weighting functions:
\begin{align}
\label{eq:weighting_functions}
    &u(z) = 1 \:\:\:\:\:\:\:\:\:\:\:\text{(non-weighted CRPS)},\\
    &u(z) = t(z,1)\:\: \text{(emphasis on the center of the distribution)},\notag\\
    & u(z) = 1-T(z,1)\: \text{(emphasis on the left tail of the distribution)},\notag\\
    & u(z) = T(z,1)\: \text{(emphasis on the right tail of the distribution)},\notag
\end{align}
where $t(z,1)$ and $T(z,1)$ are the probability density function and the cumulative density function of a t-distributed random variable with one degree of freedom, respectively. \cite{amisano2007} and \cite{gneiting2011} suggest to use the density and cumulative distribution functions of a normal random variable in the weighting function, however, in order to accommodate the heavy tailed nature of losses from cyber events, we argue that the t-distribution with 1 degree of freedom is more suitable. Typically, the CRPS can also be rewritten as \citep{gneiting2011,szekely2013, rizzo2009}:

\begin{align}
\label{eq:scoring}
    CRPS(F,y) &=   \int_{-\infty}^{\infty} \left(F(z) - \mathbb{I}_{y\leq z} \right)^2dz,\notag\\
    & =  \int_{-\infty}^\infty  F(z)\left(1-\mathbb{I}_{y\leq z}\right)dz  + \int_{-\infty}^\infty \mathbb{I}_{y\leq z}\left(1-F(z)\right)dz -\int_{-\infty}^\infty  F(z)\left(1- F(z)\right)dz \notag\\
    & = \mathbb{E}\left[|X - y|\right] -\dfrac{1}{2}\mathbb{E}\left[|X - X'|\right], 
\end{align}
where $X$ and $X'$ are two independent copies of a random variable with distribution $F$. The formulation in Equation (\ref{eq:scoring}) allows to distinguish between two components of scoring rules: reliability and uncertainty. The reliability component is captured by the first expectation, and it measures how closely the distribution F can approximate the sample realization $y$. The uncertainty component is expressed by the second expectation, and it measures how much uncertainty is carried in the distribution $F$. 
As a generalization of the CRPS, \cite{gneiting2007B} introduced the Energy Score (ES), extending the scoring rules to non linear moments:
\begin{equation}
    ES(F,y) = \mathbb{E}\left[|X - y|^{\beta}\right] -\dfrac{1}{2}\mathbb{E}\left[|X - X'|^{\beta}\right]
\end{equation}
for $\beta\in(0,2)$ such that $\dfrac{1}{2}\mathbb{E}\left[|X |^{\beta}\right]<\infty$. When $\beta=1$, the ES is equivalent to CRPS, when $\beta=2$ it reduces to the negative squared error. 

Assessing the forecasting ability of competing cyber risk classifications can be achieved by setting up a rolling window. We consider a rolling window of five years, with one year step size: at each time step, we estimate the model using the maximum likelihood in Equation (\ref{eq:likelihood}) and obtain fitted distributional parameters depending on the set of covariates $X_{i,t}$.
Let $\widehat{g}_{i,T+1|T}^1 = g\left(Y_{i,T+1};\widehat{\mu}^1_{i,T+1|T},\widehat{\tau}_{i,T+1|T}^1\right)$ and $\widehat{g}_{i,T+1|T}^2 = g\left(Y_{i,T+1};\widehat{\mu}^2_{i,T+1|T},\widehat{\tau}^2_{i,T+1|T} \right)$ be distributional density forecasts for the out of sample realization $Y_{i,T+1}$ based on two different cyber risk classifications, and $\widehat{G}^1_{i,T+1|T}$ and $\widehat{G}^2_{i,T+1|T}$ the corresponding cumulative density forecasts, where:
\begin{align}
& \log\left(\widehat{\mu}^j_{i,T+1|T}\right) = \widehat{f}_{\mu,T}(X_{i,T+1}^j) + \widehat{h}_{\mu}(T+1),\notag\\
& \log\left(\widehat{\tau}^j_{i,T+1|T}\right) = \widehat{f}_{\tau,T}(X_{i,T+1}^j) + \widehat{h}_{\tau}(T+1),\notag
\end{align}
$\widehat{f}_{\mu,T}$ and $\widehat{f}_{\tau,T}$ are the estimated linear functions based on the information up to time T, $j$ refers to the classification, and $i = 1,\dots,N_{T+1}$.
Then we can evaluate the performance of the distributional forecast via the following test of hypothesis:
\begin{equation}
\label{eq:test}
    H_0:\:\mathbb{E}\left[S(G^1,.)\right] = \mathbb{E}\left[S(G^2,.)\right]\:\: \textbf{vs}\:\: H_1:\:\mathbb{E}\left[S(G^1,.)\right] > \mathbb{E}\left[S(G^2,.)\right],
\end{equation}
where $G^1$ and $G^2$ are cumulative density distributions.
Following \cite{amisano2007}, the test statistic $T_{N_{T+1}}$ for the hypothesis test in Equation (\ref{eq:test}) is given by: 
\begin{equation}
\label{eq:test_statistic}
    T_{N_{T+1}} = \sqrt{N_{T+1}} \dfrac{ S_{N_{T+1}}^1- S_{N_{T+1}}^2}{\sigma_{N_{T+1}}},
\end{equation}
where
\begin{equation}
    S_{N_{T+1}}^1  = \dfrac{1}{{N_{T+1}}}\sum_{i=1}^{N_{T+1}} S(\widehat{G}_{i,T+1|T}^1,Y_{t+1,i}), \text{ and } S_{N_{T+1}}^2  = \dfrac{1}{{N_{T+1}}}\sum_{i=1}^{N_{T+1}} S(\widehat{G}_{i,T+1|T}^2,Y_{t+1,i})\notag
\end{equation}
are the average scores of each competing model, and
\begin{equation}
    \sigma_{N_{T+1}}^2 = \dfrac{1}{N_{T+1}} \sum_{i=1}^{N_{T+1}} (S(\widehat{G}_{i,T+1|T}^1,Y_{i,T+1}) -S(\widehat{G}_{i,T+1|T}^2,Y_{i,T+1}) )^2.\notag
\end{equation}
As shown in \cite{amisano2007} and \cite{gneiting2011}, $T_{N_{T+1}}$ is asymptotically $N(0,1)$-distributed.

The requirement of $\mathbb{E}\left[|X |^{\beta}\right]<\infty$ might be too restrictive in the context of cyber risk severity distribution, where typically the first moment is not finite \cite[see][]{eling2019,malavasi2022,shevchenko2023}. The estimates of $\tau$, obtained via the combined POT-GAMLSS approach, can be used to infer a range of admissible values for $\beta$. Nevertheless, such an admissible range might be too restrictive in the context of cyber risk severity distributions, where the penalty imposed by the non linear moment might be too harsh. For this reason, we combine the CRPS and the ES with standardized residuals of the GAMLSS approach and introduce the residual ES and CRSP:

\begin{align}
\label{eq:res_scoring}
    &rCRPS(F,y) = \mathbb{E}\left[|Z - \Phi^{-1}(F(y))|\right] -\dfrac{1}{2}\mathbb{E}\left[|Z - Z'|\right],\\
    &rES(F,y) =\mathbb{E}\left[|Z - \Phi^{-1}(F(y))|^{\beta}\right] - \dfrac{1}{2}\mathbb{E}\left[|Z - Z'|^{\beta}\right],\notag
\end{align}
where $Z$ and $Z'$ are independent copies of a standard normal distributed random variable \footnote{Alternative distributions may be considered as well in the formulation of the rCRPS and rES. One only needs to assure that the requirement $\mathbb{E}\left[|Z |^{\beta}\right]<\infty$ is satisfied. See \ref{appendix:rcrps_es_lognormal} in the online supplementary material for empirical results based on residual scoring rules defined using a lognormal distribution.}.
By using rCRPS and rES one first transforms the distributional forecasts into standardized forecasted residuals, using the inverse of the normal cumulative distribution function $\Phi^{-1}$, and then measures how close these standardized residuals are to a normal random variable. 
The interpretation of these newly introduced scores remains unchanged: a model with a higher rCRPS or rES is preferable over a model that returns low values in the forecasting exercise.

\section{Cyber Risk Classifications}
\label{sec:class}
Operational risk classifies loss events according to business lines and business sectors\footnote{Given that the vast majority of the events affect companies residing in the United States, we adopt the North American Industry Classification System (NAICS), which comprises of 20 business sectors.}. The Bank for International Settlements defines seven official event types for banking Operational Risk practices in the Basel II Accord, based on all areas of OpRisk losses. However, given the multidimensional heterogeneity that characterizes cyber events, classifying cyber risk is a complex task to the point that a world wide accepted cyber risk classification does not exist to date \cite[see, e.g.][]{peters2017}. There exist several different classifications, each one designed for specific purpose and objectives. We consider four macro-categories of cyber risk classifications: event based, operational risk based, impact-based, and based on the goodness-of-fit \footnote{\ref{appendix:mapping_classifications} in the online supplementary material presents the mapping between different cyber risk classifications.}. Although these macro-categories are non-exhaustive of all the possible cyber risk classifications, all the commonly used cyber risk classification can be allocated into one (or more) of these groups and share many similarities with these categories. Event based classifications allocate each cyber event to the main risk type defined by the underlying threat. In this category, we consider the Advisen classification which consists of 14 risk types, and the Romanosky classifications comprising of four risk types, see, e.g. \cite{romanosky2016,romanosky2019content,malavasi2022,shevchenko2023,peters2023}.
 Operational risk based classifications follow \cite{cebula2010} and \cite{cebula2014}, where cyber risk is defined as ``operational risks to information and technology assets that have consequences affecting the confidentiality, availability or integrity of information or information system". Cyber risk events are then allocated to the subcategories of operational risk: actions of people, systems and technology failures, failed internal process, and external events \citep{eling2019, eling2020cyber, eling2021}. Using the operational risk subcategories allows for a clear data identification and provides a justification for the usage of common operational risk management methodologies such as the loss distribution approach and Basel II capital requirement calculations \citep{eling2019, malavasi2022}. 
 
 Cyber events can also be classified following an impact-based approach. A common approach is to map cyber risk event frequency and severity into a risk matrix and then associate each cyber event with the corresponding label. Following the National Institute of Standards and Technology (NIST), cyber events and threats are classified via a risk matrix in terms of frequency (low, medium, high) and severity (low, medium, high) \citep{pub2004standards}. Similarly, the Australian Cyber Security Center (ACSC) applies a modified risk matrix, in terms of types of event and actors affected by the cyber threats \citep{acsc2020}. For this classification, we consider three categories of events (Low Level, Exfiltration, and Disruption) and three classes of actors affected ranked according to their importance (Low, Medium, High).
 Impact-based cyber risk classifications have the advantage of allowing the mapping of risk type and cyber threats in a more dynamic way, where for each relevant period a new risk matrix can be used to allocate cyber risk types. We take into account both the approaches of the NIST and the ACSC and consider two classifications based on risk matrices: Frequency \& Severity (based on the NIST), and Type \& Importance (based on the ACSC).

 Along these classifications, it could be useful to classify cyber events according to how good a chosen model fits the data. In particular this approach could provide interesting insights on the degree of confidence risk managers and risk officers have in their modeling approach. By grouping together events with similar model residual distributions, one could identify problematic cases and establish best practices. Given that our model produces distributional fits, we consider the standardised residual of the GAMLSS approach. One classification is constructed by fitting cyber risk related losses on the combined POT-GAMLSS approach, using the following link functions:
\begin{align}
\label{eq:link_model_selection}
    & \log(\mu(RT) )= \beta_0+\sum_{s=1}^{S} \beta_s RT_s,\\
    & \log(\tau(RT) )= \beta_0+\sum_{s=1}^{S} \beta_s RT_s,\notag
\end{align}
where $RT_s$, $s=1,\dots,S$ are dummy variables corresponding to the Advisen risk types. We then compute the normalized residual $r_{i,t}=\Phi^{-1}\left( G(y_{i,t};\hat{\mu}_{i,t},\hat{\tau}_{i,t})\right)$, with $i=1\dots,N_t$, $t=1,\dots,T$, $\Phi^{-1}$ being the inverse cumulative distribution of a standard normal random variable, and $G(.;\hat{\mu}_{i,t},\hat{\tau}_{i,t})$ being the cumulative distribution function of a generalized Pareto random variable. Finally, we merge the cyber risk types in the minimum number of categories possible, according to the relative Kolmogorov-Smirnov distance between the normalized residuals\footnote{As a robustness check we have also considered merging cyber risk types according to the Cramer-von Mises test \cite[see, e.g.][]{braun1980,csorgHo1996}. This approach provided very similar results.}. We call this classification \textit{Tail} classification. In a similar way, the \textit{Body} classification is constructed, assuming cyber risk losses follow a lognormal distribution, an assumption that is often implemented in industry applications.

\section{Data Description}
\label{sec:data}
We consider cyber event related losses in the Advisen dataset from 2008 to 2021, with a total of 165,545 observations and 62,531 entities affected. The Advisen data set is regarded as the gold standard in the cyber risk literature and has been extensively adopted in various settings \cite[see, e.g.][]{romanosky2016, aldasoro2020,aldasoro2022, malavasi2022, peters2023, shevchenko2023}. Advisen's definition of cyber risk is broader than the one used by the Privacy Clearing House, and more closely related to an Operational Risk approach \cite[see,][]{cebula2010, maillart2010, cebula2014, edwards2016, eling2019, malavasi2022}. Since we focus on the ability of cyber risk classifications to forecast the  severity distribution of future cyber events, we remove all the observations with a non-positive recorded loss \citep{edwards2016,eling2017,malavasi2022}.

Table~\ref{table:descriptive1} shows the descriptive statistics of non-zero losses for each cyber risk category based on the Advisen, Romanosky and Eling classification. Grouping the data according to the three classifications returns quite different results in terms of descriptive statistics. Nevertheless, all the risk types show high values of kurtosis, providing evidence for extreme events in the data.
\begin{table}[H]
	\centering
	\caption[Descriptive Statistics1]{This table reports some descriptive statistics of cyber risk related losses aggregated by categories, for the Adivsen, Eling and Romanosky classification, expressed in million of dollars. Even though all cyber risk types return quite different sample statistics, they all share high values of kurtosis.}
	\label{table:descriptive1}
	\begin{tabular}{l|rrrrrr}
 \toprule
		Risk Type                        &	N	& Mean	&  Median &  St. Dev. &  Skew	 & Kurt \\
		\midrule
            \multicolumn{7}{l}{Advisen Classification}\\
            Privacy - Unauthorized Contact or Disclosure	&1523	&3.59	&0.03	&26.19	&29.56	&1012.27\\
            Data - Unintentional Disclosure	                &190	&0.88	&0.11	&3.18	&6.89	&54.51\\
            Privacy - Unauthorized Data Collection	        &153	&50.38	&0.73	&413.22	&11.31	&131.68\\
            Data - Malicious Breach	                        &963	&19.05	&0.50	&161.13	&18.69	&415\\
            Identity - Fraudulent Use/Account Access	    &630	&1.10	&0.03	&6.52	&11.02	&137.75\\
            Data - Physically Lost or Stolen	            &94	    &23.4	&0.22	&206.2	&9.37	&86.82\\
            Skimming, Physical Tampering	                &86	    &1.81	&0.06	&6.28	&5.86	&39.08\\
            IT - Processing Errors	                        &45	    &70.65	&0.85	&264.43	&5.29	&29.17\\
            Phishing, Spoofing, Social Engineering	        &203	&8.11	&0.54	&51.02	&12.9	&174.18\\
            IT - Configuration/Implementation Errors	    &56	    &18.25	&0.62	&45.66	&3.01	&8.86\\
            Network/Website Disruption	                    &195	&21.68	&0.32	&67.97	&4.53	&22.02\\
            Cyber Extortion	                                &133	&1.56	&0.03	&6.37	&5.42	&30.90\\
            Industrial Controls \& Operations	            &6	    &30.7	&2.07	&68.35	&1.35	&-0.1\\
		\midrule
            \multicolumn{7}{l}{Romanosky Classification}\\

            Privacy	                                        &1676	&7.86	&0.05	&127.67	&36.18	&1395.1\\
            Data Breach	                                    &1877	&11.4	&0.19	&124.58	&23.07	&636.79\\
            Phishing and Skimming	                        &289	&6.24	&0.36	&42.96	&15.31	&247.21\\
            Security Incident	                            &435	&20.28	&0.20	&99.29	&12.14	&189.1\\
            \midrule
            \multicolumn{7}{l}{Eling Classification (operational risk based)}\\
    
            System	                                       &2306	&6.01	&0.04	&108.93	&42.38	&1917.43\\
            People	                                       &1669	&13.61	&0.30	&133.15	&21.28	&547.46\\
            Internal	                                   &302	    &28.52	&0.46	&118.21	&10.15	&131.86\\
            \midrule
            Other	                                       &27	    &1.12	&0.50	&2.26	&3.59	&13.57\\

	 	\bottomrule	
		
	\end{tabular}
\end{table}
Table \ref{table:descriptive2} shows the descriptive statistics of cyber risk losses aggregated according to the Tail classification and the Body classification. Recall that the Tail classification is based on the standardized residuals of a combined POT-GAMLSS approach, and it shows how the resulting standardized residuals can be grouped together into two groups. The Body classification follows the same principle but assumes a lognormal distribution for cyber event severity, and the corresponding standardized residuals can be grouped into three distinct groups. As it can be seen from the table, Tail and Body classifications have very different descriptive statistics, suggesting that the POT-GAMLSS should have distinguishable forecasting ability in comparison to a lognormal based severity model.
\begin{table}[H]
	\centering
	\caption[Descriptive Statistics]{This table reports some descriptive statistics of cyber risk related losses aggregated by categories, for Tail, and Body classifications, expressed in million of dollars. The tail classification is based on the standardized residual of a POT-GAMLSS model, while the Body classification is based on the standardized residuals of a Lognormal-GAMLSS specification. The risk type "Other" is excluded from the event allocation exercise. }
	\label{table:descriptive2}
	\begin{tabular}{l|rrrrrr}
 \toprule
		Risk Type                        &	N	& Mean	&  Median &  St. Dev. &  Skew	 & Kurt \\
		\midrule
            \multicolumn{7}{l}{Tail Classification}\\
             Type 1	&3526	&9.73	&0.13	&126.04	&29.5	&1013.42\\
             Type 2	&751	&14.19	&0.29	&80.41	&14.3	&267.21\\
            \midrule
            \multicolumn{7}{l}{Body Classification}\\
		Type 1	&2046	&7.67	&0.08	&123.71	&34.62	&1329.24\\
            Type 2	&1240	&19.09	&0.50	&150.92	&18.48	&422.30\\
            Type 3	&991	&5.39	&0.04	&32.12	&9.99	&115\\
            \bottomrule
	\end{tabular}
\end{table}
Table \ref{table:descriptive3} shows the descriptive statistics of the two impact-based classifications: Frequency \& Severity and Type \& Importance. The risk matrices used in allocating events into these classification are quite different and so are the descriptive statistics of the resulting risk types. It is worth mentioning that while the previously discussed classifications could be considered static, the risk types in the Frequency \& Severity and Type \& Importance classification can be allocated dynamically in our rolling window framework. Thus, an event that in a period of time is allocated to a risk type from this classification, might be allocated to a different risk type in a different window period. This has the advantage that the chosen approach might be able to replicate the ever changing behavior of cyber risk through time. One could expect that this type of classification will potentially provide a superior forecasting performance in comparison to the other static classifications.
\begin{table}[H]
	\centering
	\caption[Descriptive Statistics]{This table reports some descriptive statistics of cyber risk related losses aggregated by categories, for the Frequency\& Severity and Type \& Importance classifications expressed in million dollars. Frequency\& Severity is based on a frequency-severity risk matrix as suggested by \cite{pub2004standards}; Type \& Importance is based on event type-importance of the company involved in the attack risk matrix \cite{acsc2020, shevchenko2023}. The risk type ``Other" is excluded from the event allocation exercise. }
	\label{table:descriptive3}
	\begin{tabular}{l|rrrrrr}
 \toprule
		Risk Type                        &	N	& Mean	&  Median &  St. Dev. &  Skew	 & Kurt \\
		\midrule
            \multicolumn{7}{l}{Frequency \& Severity Classification}\\
            Likely-Low Severity	     &2153	&2.86	&0.03	&22.33	&33.89	&1359.50\\
            Likely-Medium Severity	 &1158	&19.49	&0.46	&149.54	&19.50	&465.50\\
            Likely-High Severity	 &203	&8.11	&0.54	&51.02	&12.90	&174.18\\
            Unlikely-Low Severity	 &323	&1.16	&0.07	&4.76	&6.68	&49.93\\
            Unlikely-Medium Severity &94	&23.40	&0.22	&206.20	&9.37	&86.82\\
            Unlikely-High Severity	 &153	&50.38	&0.73	&413.22	&11.31	&131.68\\
            Rare-Low Severity	     &86	&1.81	&0.06	&6.28	&5.86	&39.08\\
            Rare-Medium Severity	 &27	&1.12	&0.5	&2.26	&3.59	&13.57\\
            Rare-High Severity	     &107	&40.99	&0.82	&176.01	&8.03	&71.19\\
            \midrule
            \multicolumn{7}{l}{Type \& Importance Classification}\\
            Low Level-Low Importance	    &1045	&10.2	&0.02	&160.70	&28.97	&888.14\\
            Low Level-Medium Importance	    &691	&2.71	&0.24	&7.47	&5.70	&41.09\\
            Low Level-High Importance	    &333	&7.37	&0.48	&49.59	&16.75	&293.03\\
            Exfiltration-Low Importance	    &290	&26.77	&0.21	&244.60	&14.98	&237.65\\
            Exfiltration-Medium Importance	&1271	&7.95	&0.15	&77.17	&19.71	&441.39\\
            Exfiltration-High Importance	&345	&10.81	&0.23	&110.29	&17.11	&303.98\\
            Disruption-Low Importance	    &82	    &26.22	&0.30	&188.58	&8.53	&72.69\\
            Disruption-Medium Importance	&152	&19.09	&0.42	&66.69	&5.24	&29.30\\
            Disruption-High	Importance      &68	    &52.38	&2.70	&94.71	&2.26	&4.70\\
            \bottomrule
	\end{tabular}
\end{table}

\section{Empirical Results}
\label{sec:results}
In this section we present the results of the empirical analysis. The combined POT-GAMLSS approach requires two main components to be determined. First, the high enough threshold $u$ must be selected such that the distribution of the corresponding exceedances converges to a generalized Pareto distribution. We adopt the bootstrap methodology in \cite{villasenor2009}, where the appropriate threshold is selected as the lowest possible value such that a test of hypothesis of generalized Pareto distributed exceedances cannot be rejected. Table \ref{tab:thresholds} shows the threshold quantiles and values for each five years rolling window. Quantiles remain constant at around 51\%, confirming other findings in the literature assessing the threshold for cyber risk related losses at around 50\% of the sample \cite[see, e.g.][]{eling2019, malavasi2022}. Threshold values show an increasing trend through time, indicating that while an increasing number of lower sized losses is reported throughout the sample period, the extreme values recorded are more extreme. 

\begin{table}[H]
    \caption{This table shows quantiles and threshold values (in million USD) obtained via the bootstrap methodology of \cite{villasenor2009}, using a rolling window of five years. Quantiles remain consistent at around 51\% of the sample, while dollar figures show an increasing trend trough time.}
    \label{tab:thresholds}
    \centering
    \begin{tabular}{l|rr}
         \toprule
         Window& Quantile & Value ($u$)   \\
         \midrule
          2008-2012	&0.51	&0.065\\
          2009-2013	&0.51	&0.131\\
          2010-2014	&0.52	&0.189\\
          2011-2015	&0.51	&0.192\\
          2012-2016	&0.52	&0.237\\
          2013-2017	&0.51	&0.249\\
          2014-2018	&0.51	&0.254\\
          2015-2019	&0.51	&0.262\\
          2016-2020	&0.51	&0.340\\
          2017-2021	&0.51	&0.456 \\
          \bottomrule
    \end{tabular}
\end{table}

Second, the GAMLSS approach allows us to use covariates in the estimation of the parameters for the generalized Pareto distribution. Covariates are usually chosen according to problem specific features, and consistently with the literature. Following \cite{malavasi2022}, we consider log-link functions in the generalized Pareto specification and a series of company and cyber event specific covariates. To capture how cyber risk evolves through the sample period, we consider time as a covariate \citep{maillart2010, biener2015, eling2019, malavasi2022}. The dependence on time is included via cubic splines, where the optimal number of knots is decided via Akaike information criteria \cite{chavez2016, eling2019, malavasi2022}. To proxy for company size and human error, we consider two categorical variables, based on the number of employees and revenue \citep{cope2008,ganegoda2013,evans2016, eling2019}. Business sector is also used as a proxy for capturing unobservable differences between companies operating in different markets \citep{dahen2010, ganegoda2013,chavez2016,eling2019,malavasi2022,peters2023}\footnote{We consider the following business sectors:  Finance and Insurance; Administrative and Support and Waste Management and Remediation Services; Information; Professional, Scientific, and Technical Services; Public Administration; Healthcare and Social Assistance.}. Since more than 80\% of the companies affected by cyber events have their headquarters in the US, we include also location as a  dummy variable (US companies vs non-US companies) to account for regulatory differences between the US and the rest of the world \cite[see,][]{malavasi2022,shevchenko2023}. Finally, we also consider contagion as a categorical variable with three mutually exclusive events: events related to other events in the same company, events related to an event in other companies, and one shot events. 

The aforementioned covariates represent the base model and aim to capture the dependence between the cyber severity distribution, and company specific characteristics. To test for better forecasting performance of cyber risk classifications, we also include categorical variables for each risk type.  In other words, each model is compared based on the same set of company and event-specific covariates, with the only difference being the inclusion of risk types that vary with each classification. Additionally, to facilitate comparison, we have also included two additional models. One model ignores risk types and does not consider them in the GAMLSS regression, while another model involves a random classification where four categories of risk types are randomly allocated to the cyber event. In the following subsection, we compare the forecasting ability of each cyber risk classification with the model with no classification and the model with random classification via the test of hypothesis in Equation (\ref{eq:test}).

\subsection{Severity analysis of cyber risk classifications}
\begin{figure}[H]
    \centering
    \includegraphics[width = 0.9\textwidth, height = 0.7\columnwidth]{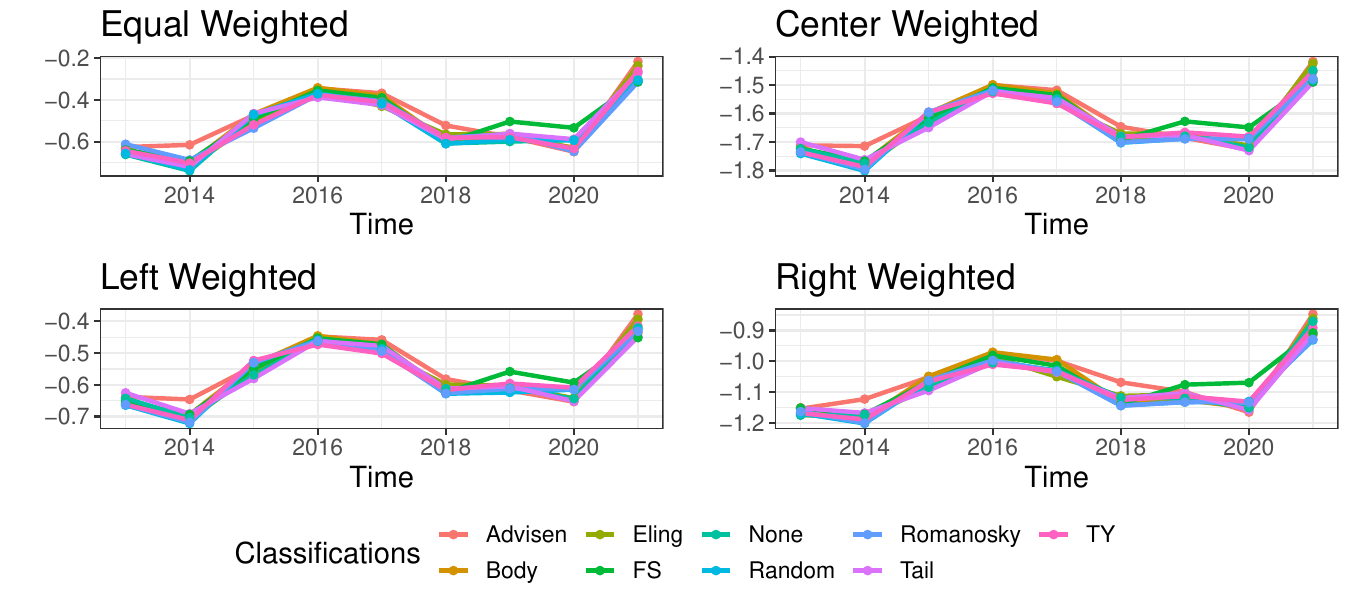}
    \caption{This figure shows the yearly averages of rCRPS for different classification for four different weighting functions on the log-scale. The Random and None classifications are included for comparison. No clear winner among the considered classifications can be identified, and often the None and Random classifications yield better distributional forecasts for future losses from cyber events in comparison to the commonly used classifications. FS and TY refer to the Frequency \& Severity and Type \& Importance classification, respectively.}
    \label{fig:crps_outsample}
\end{figure}
Figure~\ref{fig:crps_outsample} depicts the yearly averages of the rCRPS for the out of sample period, for different classifications, according to different weighting functions. The equally weighted score and the score that puts more weight on the right tail of the distribution (`right weighted') exhibit similar behavior through time, remaining relatively flat until 2020. At the same time, the scores that put emphasis on the left tail (`left weighted') and the center (`center weighted') of the distribution show an increase up to 2016, followed by a decreasing trend until 2020 and an increasing trend in 2021. It should be mentioned that in the last two years of the analysis, the number of out of sample observations are considerably lower than for the previous years, and therefore, the results for 2020 and 2021 should be considered with care, due to a possible lack of statistical power. Moreover, none of the considered classifications consistently outperforms the other classifications, i.e. none of the considered classifications seems to provide yearly average scores that are higher than the scores of any other classification. More surprisingly, almost all classifications exhibit years where their forecasting ability is either lower or indistinguishable from one of the models with a random classification or no classification. This first result seems to cast doubts on the usage of cyber risk classifications in modelling cyber event severity distributions with the intent of forecasting. 

To formally evaluate the out of sample forecasting performance of different classifications, we perform the test of hypothesis in Equation (\ref{eq:test}). Figure~\ref{fig:crps_outsample} shows that it's not possible to identify a classification with an overall better forecasting performance. Therefore, we proceed with a pairwise comparison between the out of sample performance of each cyber risk classification and the model that uses a random classification or no classification at all, respectively. Table~\ref{tab:crps_ts_outsample} shows the values of the rCRPS based test statistic for the overall out of sample period with the four weighting schemes. The first panel reports the results in the case where the forecasting performance is compared with the one from the model with no classification. The Type \& Importance classification is the only case, where the null hypothesis is rejected for all of the four weighting schemes. This classification requires a higher level of sophistication, as it dynamically changes through time and it allocates the riskiness across the actors involved in the cyber breaches, whose importance is adjusted trough time \citep{acsc2020}. Nonetheless, the evidence of better performance of the Type \& Importance classification could be considered weak, as the null is only rejected at the 5\% significance level and not at the 1\% level. Since our aim is to better forecast the severity distribution of cyber events, particular focus should be given to the right tail weighted test statistic. In this case, the null hypothesis is rejected also for the Romanosky classification, indicating that this event based classification can have better performance than a model without any cyber risk classification. Although, the null is rejected only at the 5\% level of significance. 

Similar considerations can be made by looking at the bottom panel of Table~\ref{tab:crps_ts_outsample}, where each classification forecasting ability is compared with a model based on a random classification. In this case, the Type \& Importance and the Romanosky classification outperform the random classification model for almost all weighting schemes. However, when examining the right tail weighted scores, the null hypothesis is rejected only for the Romanosky and Tail (at the 5\% significance level) classifications. As previously discussed, the right tail weighted test statistic is arguably the most important one, since extra focus is posed on the part of the distribution where extreme and catastrophic events are generated. Failing to reject the null in this instance is symptomatic of the inadequacy of cyber risk classifications in the context of forecasting the severity distribution of future cyber events.

\begin{table}[H]
    \caption{This table shows the values of the test statistic based on rCRPS for the overall out of sample period. The critical values are 1.64 (5\% level of significance) and  2.32 (1\% level of significance).}
    \label{tab:crps_ts_outsample}
    \centering
    \begin{tabular}{l|rrrr}
    \toprule
    \multicolumn{5}{c}{Classification vs None Classification }\\
Classifications	&Equal&	Center&	Left	&Right\\
\midrule
Advisen	&1.46	&1.45	&1.45	&1.47\\
Romanosky	&1.70&	1.56&	1.57&	2.28\\ 
Eling	&1.51&	1.49&	1.49&	1.55\\
Tail	&1.49 &	1.45&	1.46&	1.54\\
Body	&1.48&	1.45&	1.45&	1.51\\
Frequency \& Severity	&1.43&	1.43	&1.43&	1.44\\
Type \& Importance	&2.07&	1.91&	2.01&	1.81\\

\midrule
\multicolumn{5}{c}{Classification vs Random Classification }\\
Classifications	&Equal&	Center&	Left	&Right\\
\midrule
Advisen &	1.46&	1.45	&1.46	&1.47\\
Romanosky	&2.06	&1.69	&1.76	&4.93\\
Eling	&1.57	&1.56	&1.56	&1.60\\
Tail	&1.67	&1.56	&1.59	&1.74\\
Body	&1.51	&1.47	&1.47	&1.55\\
Frequency \& Severity	&1.44	&1.44&	1.44	&1.44\\
Type \& Importance	&3.74&	3.26	&7.86	&1.15\\
\bottomrule
\end{tabular}
\end{table}
Table~\ref{tab:proportion} shows the proportion of times, over the nine years out of sample period, the null hypothesis is rejected for different classifications and different weighting functions in the rCRPS case (based on conducting a separate test each year). Looking at both the upper and lower panels, the Advisen classification seems to perform better in comparison to the other classifications, with a rejection rate of more than 50\% and more than 60\% with respect to the none classification and the random classification models, respectively. However, it is worth noting that the sample size used in computing the test statistic on a yearly basis might not be large enough to guarantee convergence to a normal distribution. Therefore, results in Table~\ref{tab:proportion} should be interpreted with caution given the lack of power of the test statistic in small sample sizes.

\begin{table}[H]
    \caption{Proportion of times, over the nine years out of sample period, the null hypothesis is rejected for the rCRPS based test at the 5\% significance level.}
        \label{tab:proportion}
    \centering
    \begin{tabular}{l|cccc}
        \toprule

    \multicolumn{5}{c}{Classification vs None Classification }\\
    Classifications	&Equal&	Center&	Left	&Right\\
    \midrule
    Advisen & 0.33 & 0.55 & 0.55 & 0.55\\
    Romanosky & 0.33 & 0.11 & 0.11 & 0.44 \\
    Eling & 0.33 & 0.33 & 0.33 & 0.33 \\
    Tail& 0.33 & 0.11& 0.22 & 0.33\\
   Body & 0.00    & 0.22 & 0.44 & 0.11 \\
   Frequency \& Severity & 0.22 & 0.22 & 0.33 & 0.44 \\
    Type \& Importance& 0.11 & 0.22 & 0.11 & 0.22 \\
    \midrule
    \multicolumn{5}{c}{Classification vs Random Classification }\\
    Classifications	&Equal&	Center&	Left	&Right\\
    \midrule
    Classifications	&Equal&	Center&	Left	&Right\\
    Advisen & 0.33& 0.44 & 0.44 & 0.66\\
    Romanosky &0.33 & 0.22& 0.33 &0.44\\
    Eling&0.44 &0.44 & 0.44 & 0.33\\
    Tail&0.33 & 0.11 &0.44& 0.22\\
    Body&0.11 & 0.22 & 0.33 & 0.11\\
    Frequency \& Severity&0.55 &0.33 & 0.33 & 0.33\\
    Type \& Importance&0.33 & 0.22 &0.22 &0.22\\
\midrule    
    \end{tabular}
\end{table}

\begin{figure}[H]
    \centering
    \includegraphics[width = 0.9\textwidth, height = 0.7\columnwidth]{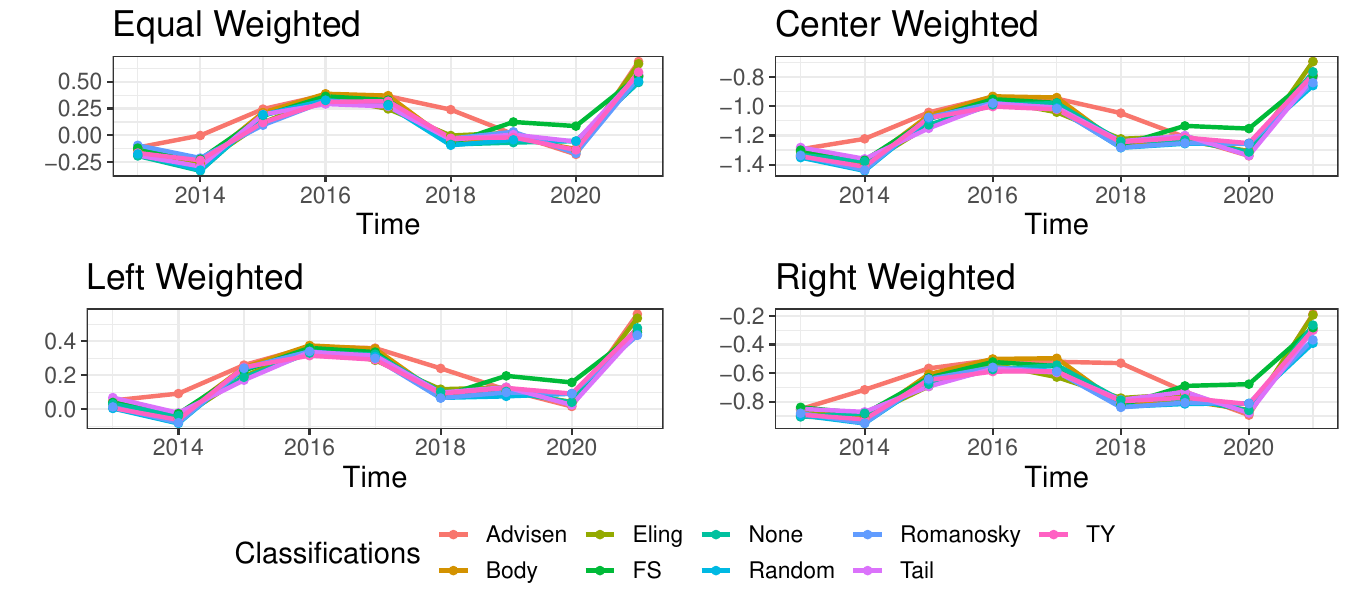}
    \caption{This figure shows the yearly averages of rES for the different classifications for four different weighting functions on the log-scale. The Random and None classifications are included for comparison. No clear winner among the considered classifications can be identified, and often the None and Random classifications yield better results in forecasting the severity distribution of future cyber events than commonly used classifications. }
    \label{fig:es_outsample}
\end{figure}

Similarly to the rCRPS case, Figure \ref{fig:es_outsample} shows the yearly average values of rES for different classifications. In this case, the scores have a similar behavior through time, with the change in the weighting function seemingly affecting only the score magnitude rather than the overall performance. In terms of better performance, it seems that all the cyber risk classifications have a a very similar performance during the years 2015-2018, with the Advisen classification indicating a marginally better forecasting ability.

Table \ref{tab:es_testStats} reports the values of the rES test based statistics for the overall out of sample period. Similarly to the rCRPS case, when the forecasting ability is compared against the performance of a model without any cyber risk classification, the null hypothesis is rejected only for the Type \& Importance classification. All other classifications exhibit a  forecasting performance that is statistically indistinguishable from the one of a model where no classification is used. When cyber risk classifications are compared to a model based on a random classification, the null hypothesis is rejected for the Type \& Importance, Tail, and Romanosky classifications, although only at the 5\% level of significance.

\begin{table}[H]
    \caption{This table shows the values of the test statistic based on rES for the overall out of sample period. The critical values are 1.64 (0.95\%) and  2.32 (0.99\%).}
 \label{tab:es_testStats}    
    \centering
    \begin{tabular}{l|rrrr}
    \toprule
    \multicolumn{5}{c}{Classification vs None Classification }\\
Classifications	&Equal&	Center&	Left	&Right\\
\midrule
Advisen&	1.51&	1.49	&1.49	&1.51\\
Romanosky	&1.57	&1.53	&1.52	&1.58\\ 
Eling	&1.51&	1.51	&1.51	&1.52\\
Tail	&1.51	&1.48&	1.49	&1.53\\
Body	&1.47	&1.45	&1.45&	1.47\\
Frequency \& Severity	&1.44	&1.44	&1.44	&1.44\\
Type \& Importance	&1.69	&1.88	&1.84	&1.74\\
\midrule
    \multicolumn{5}{c}{Classification vs Random Classification }\\
Classifications	&Equal&	Center&	Left	&Right\\
\midrule
Advisen &	1.51&	1.50	&1.50&	1.52\\
Romanosky	&1.76&	1.62	&1.63	&1.75\\
Eling	&1.60&	1.60	&1.60	&1.61\\
Tail&	2.11&	1.73	&1.76	&2.18\\

Body	&1.49&	1.46&	1.47&	1.51\\
Frequency \& Severity&	1.45	&1.45	&1.45&	1.45\\
Type \& Importance	&1.93&	2.62	&2.46	&1.98\\

\bottomrule
    \end{tabular}
\end{table}

Table \ref{tab:proportion_es} reports the proportion of yearly scores for which the null hypothesis is rejected. Similarly to the rCRPS case, on a yearly basis, the Advisen classification performs slightly better than the other classifications considered. Nonetheless, overall these results confirm our previous findings: severity modeling based on cyber risk classifications produces forecasts that are often not statistically distinguishable from a model where the risk type component is disregarded. When the forecasting ability is compared with the one of a model based on a random classification, the improvements tend to be minimal and many classifications do not provide an out of sample performance that is distinguishable from a model based on a random allocation of risk types.

\begin{table}[H]
    \caption{Proportion of times, over the nine years out of sample period, the null hypothesis is rejected for the rES test at the \% significance level.}
    \label{tab:proportion_es}
    \centering
    \begin{tabular}{l|cccc}
    \toprule
    \multicolumn{5}{c}{Classification vs None Classification }\\
Classifications	&Equal&	Center&	Left	&Right\\
\midrule
Advisen& 0.44 & 0.55 & 0.55 & 0.55\\
Romanosky& 0.55 & 0.22 & 0.33 & 0.33\\
Eling& 0.33 & 0.22 & 0.22 & 0.33\\

Tail& 0.33 & 0.22 & 0.44 & 0.33\\

Body& 0.11 & 0.44 & 0.44 & 0.11\\
Frequency \& Severity& 0.44 & 0.33 & 0.33 & 0.44\\
Type \& Importance& 0.33 & 0.33 & 0.33 & 0.33\\
\midrule
    \multicolumn{5}{c}{Classification vs Random Classification }\\
Classifications	&Equal&	Center&	Left	&Right\\
Advisen& 0.55& 0.55 & 0.44 & 0.44\\
Romanosky &0.33 & 0.33 & 0.22 & 0.33\\
Eling&0.22 & 0.44 & 0.55 & 0.33\\
Tail&0.22 & 0.22 & 0.44 & 0.22\\

Body&0.11 & 0.55 & 0.44 & 0.11\\
Frequency \& Severity&0.55 & 0.33 & 0.33 & 0.44\\
Type \& Importance& 0.33 & 0.33 & 0.22 & 0.33\\
\midrule
    \end{tabular}
\end{table}

\begin{figure}[H]
    \centering
    \includegraphics[width = 0.9\textwidth, height = 0.4\columnwidth]{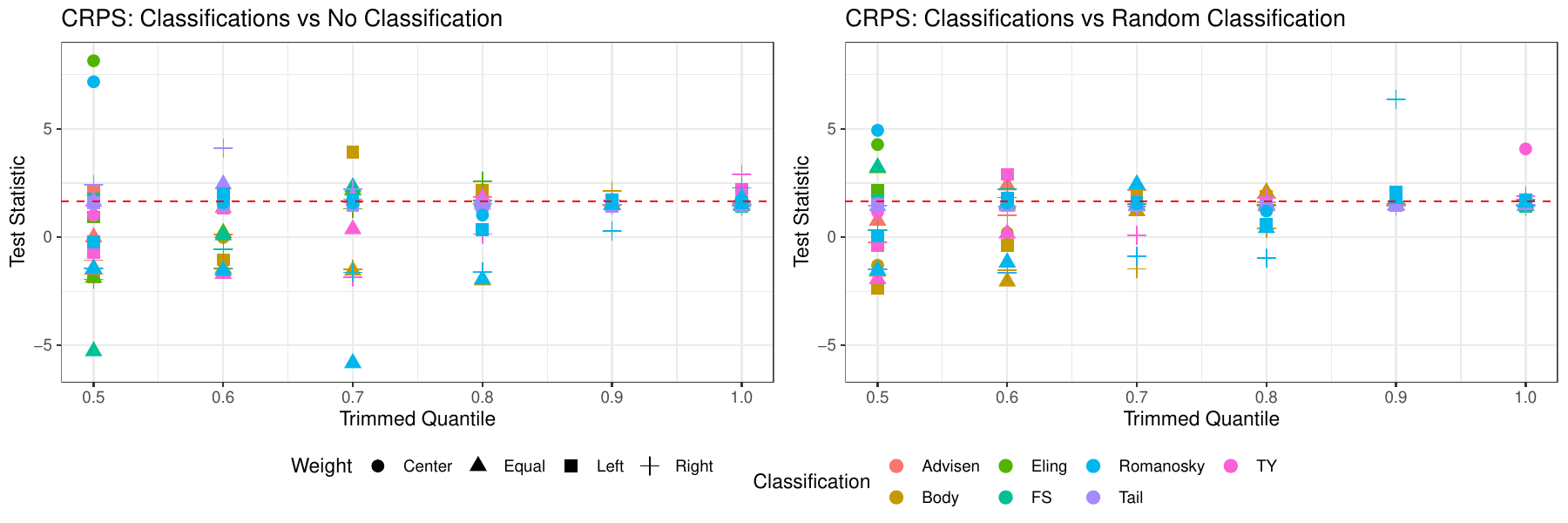}
    \caption{This figure shows the test statistics corresponding to rCRPS for different classifications, under different weighting schemes. The trimmed quantiles are 50\%, 60\%, 70\%, 80\%,and 90\%. For completeness, the test statistic corresponding to the full sample is included as well. The red dashed line corresponds to the critical values for the one sided test at the 5\% level of significance. }
    \label{fig:trimmed_crps}
\end{figure}

\begin{figure}[H]
    \centering
    \includegraphics[width = 0.9\textwidth, height = 0.4\columnwidth]{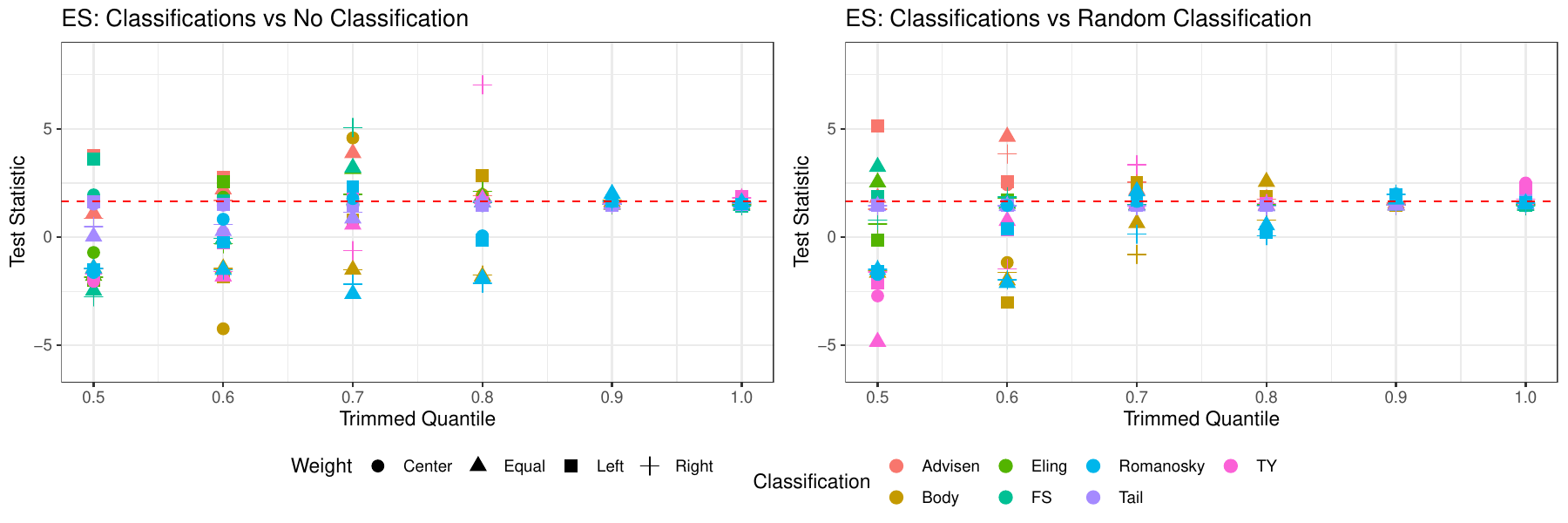}
    \caption{This figure shows the test statistics corresponding to rES for different classifications, under different weighting schemes. The trimmed quantiles are 50\%, 60\%, 70\%, 80\%,and 90\%. For completeness, the test statistic corresponding to the full sample is included as well. The red dashed line corresponds to the critical values for the one sided test at the 5\% level of significance.}
    \label{fig:trimmed_es}
\end{figure}

Before concluding this section, we perform an additional robustness check aimed at verifying whether the results of the our hypothesis tests are affected by the presence of extreme events in the data. To control and evaluate the impact of heavy tails on our methodology, we perform the test of hypothesis in (\ref{eq:test}) using a trimmed cyber loss distribution. We trim the out of sample losses at deciles, starting from the median and then consecutively increase the sample size until the full sample is included in the test. The motivation behind this robustness check is in the spirit of \cite{peters2023}: if the poor out of sample performance of the estimated severity distribution for the considered classifications was due to the presence of very extreme losses that are almost impossible to forecast, then removing these losses should improve the overall forecasting ability of the classifications. Then one may conclude that using cyber risk classifications in severity modeling still carries some meaningful improvements, even though the very extreme losses remain almost impossible to forecast. Figures \ref{fig:trimmed_crps} and \ref{fig:trimmed_es} show the results of test of hypothesis on the trimmed samples, where the red dashed horizontal line corresponds to the 5\% critical values that will lead to a rejection of the null. 

As it can be seen by both figures, trimming the sample does not increase the out of sample performance of the models. To the contrary, trimming often seems to generate the opposite effect, such that in many cases it is not possible to reject the null hypothesis. Moreover, as the trimmed quantile increases the values of the test statistic increases, increasing the chances of rejecting the null. Although this results may seem surprising, it indicates that the combined POT-GAMLSS approach works relatively well in capturing the tail behavior of the severity distribution of cyber events and including cyber risk classification in the modelling might only provide limited benefits, or it might even have a confounding effect of the estimates of the distributional parameters.

The failure of cyber risk classifications to generate statistically better forecasting performance has indeed important implications on the cyber risk modeling literature. Several studies have used cyber risk types corresponding to the considered cyber risk classifications and the results are mixed. \cite{romanosky2016} and \cite{aldasoro2022}, basing their analysis on Romanosky classifications, suggest that risk types are important risk drivers for the quantification of cyber risk. In view of our results, one would need to exercise extra caution in using their results for forecasting purposes, as risk types might actually behave more similarly to confounding variables rather than the applied independent variables in regression based modeling.  \cite{biener2015,eling2015modelling,eling2019} included cyber risk types in their regression framework (based on the classification suggested by Eling) and do not obtain coefficients that are statistically significant, what is consistent with out findings. Similar results are also obtained by \cite{malavasi2022}, based on the Advisen classification. Our results strongly indicate that, from a statistical point of view, cyber risk classifications potentially only provide limited benefits in modeling the severity of cyber events. In particular in terms of enhancing the forecasting performance of a model, classifications do not seem to be distinguishable from confounding variables. Nonetheless, the benefit of cyber risk types for business risk management purposes and business motivated objectives remains strong. For cyber risk business management, modeling can benefit from including cyber risk types, applying text based algorithms in threat identification, developing mitigation strategies, and cyber risk scoring \cite[see, e.g.][]{gatzert2022,shetty2018, eling2021}.

The unsatisfactory performance of the Frequency \& Severity classification in forecasting cyber risk severity is somewhat more surprising. Given the dynamic nature of cyber risk, one would expect that a dynamically updated classification will yield a better performance than a static classification. However, our results suggest that
even using well established and widespread methods such as risk matrices and impact-based classifiers to create cyber risk classifications and allocate risk types to cyber events, might not be sophisticated enough to capture the dynamic heterogeneity of cyber risk. This result needs to be interpreted jointly with the stream of literature that suggests to adopt the Frequency \& Severity classification in \cite{pub2004standards}. In particular,  \cite{eisenbach2022cyber}, \cite{healey2018future}, \cite{curti2023cyber} and \cite{duffie2019cyber} analyze how cyber risk can contribute to market risk, and show how financial institutions might be resilient enough to survive market turmoil generated by cyber risk events. However in view of our results, one should be careful in relying on forecasting based on a Frequency \& Severity classification, as the ability of this classification to forecast future cyber risk losses appears to be limited.

Third, the classification by Eling corresponds to the definition of cyber risk as a subclass of operational risk. The unsatisfactory performance of this classification clearly indicates that cyber severity modelling based on operational risk approaches might not be statistically effective from a forecasting exercise point of view. This aspect is in contrast with the standard view in the literature of considering cyber risk as a subcategory of operational risk \cite[see,][]{cebula2010,cebula2014,eling2015modelling,biener2015,eling2018,cohen2019,eling2019,gatzert2022,aldasoro2022}. When it becomes relevant to obtain reliable forecasts, our results suggest that perhaps, defining cyber risk as a subcategory of operational risk, might have confounding implications on the cyber event severity distribution. This last point is consistent with other findings in the literature suggesting that, even if transmission channels of cyber risk are similar to those of operational risk, the unique nature of cyber events required cyber risk to be treated separately from classic operational risk \citep{kashyap2019some,adelmann2020cyber}. This line of thought is further supported by the threshold values required for the POT methodology. For cyber risk modelling, the literature seems to agree that values between 50\% and 60\% of the sample are high enough in order to invoke the Belkema, De Hann, Pickands theorem, while for operational risk the threshold is usually higher than 90\% \cite[see,][]{balkema1974, pickands1975,eling2019, malavasi2022}. 

\subsection{Frequency analysis of cyber risk classifications}

Results of the previous section indicate that the forecasting ability of cyber risk classifications is limited, even when classifications are constructed dynamically. In this section we investigate whether this effect is present also in modelling the frequency of cyber events. We perform a two sample chi-squared test to verify if the frequency distribution of cyber events based on different cyber risk classifications can be distinguish from a frequency modelling based either on no classification or a random based classification. Table \ref{tab:frequency_outsample_year} shows the p-values for the conducted chi-squared test for the cyber event frequency distribution for each out of sample year. Hereby, the null hypothesis is that the model performance is not distinguishable from either the none classification (top panel) or random classification (bottom panel). We find that in all the cases considered, there is enough evidence to reject the null hypothesis of an equal performance. Therefore, modeling the frequency distribution of cyber events based on a cyber risk classification provides results that are distinguishable from using non-informative models. Repeating the same exercise for the overall sample period, returns similar conclusions and confirms that cyber risk classifications are indeed useful in estimating and modeling the frequency of cyber events. 

 Overall, we find that cyber risk classifications can be used to better fit the frequency distribution of losses from cyber events. At the same time, the classifications are typically unable to provide significantly better forecasts for the severity of cyber risks. These results have important implications for the insurance sector. Based on these findings, an insurance company might decide to disregard cyber risk types or classifications in severity modelling, while including them in modelling the frequency of cyber events. In doing so, insurance premiums derived from a unique cyber severity model can then be adjusted following cyber event frequency estimates, which depends on cyber risk types. This last aspect is often neglected in the literature on cyber risk rate making, where often classification algorithms are used \cite[see, e.g.][]{dacorogna2023managing,dacorogna2023building,farkas2021cyber}.

\begin{table}[]
    \caption{This table shows the results for a conducted chi-squared test on the difference in the out of sample forecasting performance for cyber event frequency. The frequency is modeled using the considered cyber event classifications, and either the none classification (top panel) or the random classification model (bottom panel). All cyber event classifications provide a performance superior to that of the uninformative models. }
    \label{tab:frequency_outsample_year}
    \centering
    \begin{adjustbox}{width=1\textwidth}
    \begin{tabular}{l|rrrrrrrrr}
    \toprule
    &\multicolumn{9}{c}{Classification vs None Classification }\\
    	 &2013	&2014	&2015	&2016	&2017	&2018	&2019	&2020	&2021\\
      \midrule
Advisen	                &$<$0.001& $<$0.001&	0.109& $<$0.001& $<$0.001&  0.084&	0.120&	0.199&	0.199\\
Romanosky	        &$<$0.001& $<$0.001&	0.109& $<$0.001& $<$0.001&	0.084&	0.084&	0.199&	0.199\\
Eling	                &$<$0.001& $<$0.001&	0.109& $<$0.001& $<$0.001&  0.084&	0.084&	0.199&	0.199\\
Tail	            &$<$0.001& $<$0.001&	0.109& $<$0.001& $<$0.001&	0.084&	0.084&	0.199&	0.199\\

Body	                &$<$0.001& $<$0.001&	0.109& $<$0.001& $<$0.001&	0.084&	0.084&	0.199&	0.199\\
Frequency \& Severity	&$<$0.001& $<$0.001&	0.109& $<$0.001& $<$0.001&	0.084&	0.120&	0.199&	0.199\\
Type \& Importance	    &$<$0.001&$<$0.001&	    0.109& $<$0.001& $<$0.001&	0.084&	0.084&	0.199&	0.199\\
\midrule
&\multicolumn{9}{c}{Classification vs Random Classification }\\
    	 &2013	&2014	&2015	&2016	&2017	&2018	&2019	&2020	&2021\\
\midrule
Advisen	&$<$0.001&$<$0.001&	0.051 &	$<$0.001 &$<$0.001&	0.064&	0.043&	0.199&	0.199\\
Romanosky	&$<$0.001&$<$0.001&	0.051 &$<$0.001	&$<$0.001	&0.064&	0.064&0.199&	0.199\\
Eling	&$<$0.001&$<$0.001 &	0.051& 0.001&$<$0.001 &	0.064 & 0.064& 	0.199&	0.199\\
Tail	&$<$0.001&$<$0.001	&0.051 &	$<$0.001	 &$<$0.001	 &0.064&	0.064&	0.199&	0.199\\

Body	&$<$0.001&$<$0.001 &	0.051 &	$<$0.001 & $<$0.001&	0.064& 	0.064&	0.199&	0.199\\
Frequency \& Severity	&$<$0.001&$<$0.001	&0.051 &	$<$0.001	&$<$0.001 &	0.064&	0.043&	0.199&	0.199\\
Type \& Importance	&$<$0.001&$<$0.001&	0.051	&0.001&$<$0.001 &	0.064&	0.064&	0.199&	0.199\\

      \midrule
    \end{tabular}
    \end{adjustbox}
\end{table}

\clearpage

\subsection{Power Analysis}
\label{sec:power}
In this subsection we present a robustness analysis on the statistical power of the test of hypothesis presented in the previous subsections via a simulation set up. To ensure relevance to cyber risk, we consider the cyber event severity model based on the Type \& Importance classification, as it is the classification returning arguably the best results among the considered classifications. We use the model estimated during the subsample period 2009 to 2014 to generate observations corresponding to the year 2015. For each out of sample observation, we simulate 1000 draws from $G_{i,2015|2014}^{TI}$  \footnote{The suffix $TI$ indicates the model corresponds to the parameters obtained using the Type \& Importance classification.}. The total number of sample observations is 148 in the year 2015, from which the simulated universe of cyber risk losses consists of a total of 148,000 observations. Table \ref{tab:stats_power} shows the descriptive statistics of losses due to cyber events that occurred during the out of sample period.

\begin{table}[]
    \caption{Descriptive statistics of losses due to cyber events occurred during the year 2015, broken down by cyber risk types of the Type \& Importance classification. No events of the type Low Level-Low Importance were recorded during the year 2015. }
    \label{tab:stats_power}
    \centering
    \begin{tabular}{l|rrrrrr}
    \toprule
                     Risk Type & N  & Mean &Median & St.Dev. & Skew & Kurt\\
                     \midrule

Low Level-Medium Importance   &26  &37.18   &0.68 &181.15 &4.52 &19.22\\
Low Level-High Importance     &70   &3.99   &1.50   &6.14 &2.79 &7.75\\
Exfiltration-Low Importance    &12  &17.95   &7.10  &25.71 &1.54  &1.24\\
Exfiltration-Medium Importance&11 &101.23   &0.90 &298.55 &2.45  &4.48\\
Exfiltration-High Importance  &16   &2.46   &1.62   &2.75& 1.66  &2.44\\
Disruption-Low Importance     &2   &5.12   &5.12   &6.91 &0.00 &-2.75\\
Disruption-Medium Importance   &5   &9.71   &7.50  &10.04 &0.23 &-2.05\\
Disruption-High Importance     &6   &2.93   &0.34   &6.40 &1.36 &-0.08\\
\bottomrule
    \end{tabular}
\end{table}

To evaluate the power of the test, we draw 10,000 random samples of different sizes from the generated universe of losses and perform the test of hypothesis in \eqref{eq:test}, where under the null hypothesis the out of sample performance of the Type \& Importance classification is not distinguishable from the out of sample performance of the no classification model, or a random classification model. We test the null against the alternative hypothesis of a better performance by the Type \& Importance classification. The power of the test is computed as the number of times the hypothesis is rejected.

Table \ref{tab:power_rCRPS} shows the results of the power analysis in the case of the rCRPS based test statistic with different weighting regimes. Looking at the column corresponding to the equally weighted scores, the null hypothesis is never rejected, indicating that the sample size required for the test statistic to correctly distinguish between the forecasting performances of the two models is far greater than the sample size considered. To put this into perspective with regards to cyber risk and availability of data, in our out of sample study there are a total of 1,380 observations, collected over a period of nine years. Therefore, if data continues to be collected at the same pace it has been collected until now, we would require an additional 40 to 50 years' worth of data to achieve a satisfactory level of statistical power in the case of the equally weighted rCRPS. In the context of cyber risk this is an absurdly long period, since characteristics of cyber risk tend to change dynamically and rapidly through time. Nonetheless, as it can be seen from the other columns of Table~\ref{tab:power_rCRPS}, including weighting schemes in the scores improves statistical power, even with small sample sizes.

Putting these results in the context of Figures \ref{fig:crps_outsample} and \ref{fig:es_outsample} and Tables \ref{tab:proportion} and \ref{tab:proportion_es}, where the sample size used to calculate the test statistic is approximately in the range of 100-200 observations, should reinforce the need for caution in interpreting these results. Even if in small samples the results appear to be more in favour of cyber risk classifications, the lack of power of the test should be considered before drawing any meaningful conclusions. Nonetheless, with a sample size greater than 100 the test achieves a satisfactory level of power, especially for the weighting scheme that put more emphasis on the right tail of the distribution. 
\begin{table}[H]
    \caption{This table shows the results of the power analysis in the case of using the rCRPS. For each sample size, 10,000 random draws are sampled from the generated losses from cyber events. The power of the test is computed as the fraction of random draws for which the null hypothesis of equal performance is rejected.}
    \label{tab:power_rCRPS}
    \centering
    \begin{tabular}{l|rrrr}
    \toprule
    \multicolumn{5}{c}{Type \& Importance vs None Classification }\\
        Sample Size & Equal& Left& Center & Right \\ 
            \midrule
20 &	0&	1&	0&	0.410\\
50	&0	&1&	0&	0.930\\
100	&0	&1&	0&	1\\
500	&0	&1&	0.006&	1\\
1000 & 0	&1	&0.038&	1\\
2000 & 0	&1	&0.256	&1\\
5000 & 0	&1	&0.926	&1 \\
\midrule
\multicolumn{5}{c}{Type \& Importance vs Random Classification }\\
        Sample Size & Equal& Left& Center & Right \\ 
\midrule
20 &0	&1	&0	&0.418\\
50&0	&1	&0	&0.928\\
100&0	&1	&0	&1\\
500 &0	&1	&0.014	&1\\
1000 &0	&1	&0.058	&1\\
2000 & 0&	1&	0.391&	1\\
5000 & 0&	1&	0.976&	1\\
\bottomrule
    \end{tabular}
\end{table}

Table \ref{tab:power_rES} shows the power analysis for the alternative rES. In this case, the power converges slower than for the rCRPS case, but reaches satisfactory levels at sample size greater than 500 observations both for the weighting schemes that put more emphasis on the left and right tail of the distribution.

\begin{table}[H]
    \caption{This table shows the results of the power analysis in the case of using the rES. For each sample size, 10,000 random draws are sampled from the generated losses due cyber events. The power of the test is computed as the fraction of random draws for which the null hypothesis of equal performance is rejected.}
    \label{tab:power_rES}
    \centering
    \begin{tabular}{l|rrrr}
    \toprule
    \multicolumn{5}{c}{Type \& Importance vs None Classification }\\
        Sample Size & Equal& Left& Center & Right \\ 
            \midrule
20  &0	&0.654	&0	&0.028\\
50	&0	&0.99	&0	&0.181\\
100	&0	&1	&0	&0.492\\
500	&0	&1	&0	&1\\
1000	&0	&1	&0	&1\\
2000	&0	&1	&0	&1\\
5000 &0	&1	&0	&1\\   
\midrule
\multicolumn{5}{c}{Type \& Importance vs Random Classification }\\
        Sample Size & Equal& Left& Center & Right \\ 
\midrule
20&   0 &	0.642	&0	&0.034\\
50&  0	&0.992	&0&	0.182\\
100& 0	&1	&0&	0.488\\
500& 0	&1	&0&	1\\
1000 & 0 & 1 & 0 & 1\\
2000 & 0 &1 &0 & 1\\ 
5000 & 0 &1 &0 & 1\\ 
\bottomrule
    \end{tabular}
\end{table}

\section{Conclusions}
\label{sec:con}
Identifying cyber risk exposures has become of fundamental importance for individuals, businesses, and public entities. Cyber risk classifications play a pivotal role in establishing effective cyber risk management strategies, computing insurance premiums calculations, and in developing regulatory policies and frameworks. In this study we have analyzed the out of sample forecasting performance of different approaches to cyber risk classification. Our analysis involved a comprehensive investigation of various classifications, including event-based, operational risk-based, and risk matrix-based types \citep{pub2004standards, cebula2010, cebula2014, romanosky2016, eling2019, nai2018, nai2019, acsc2020}. To tackle the complex and dynamic nature of cyber risk, we have used a combined approach of  dynamic extreme value theory methods and generalized models models for location scale and shape regression framework \citep{rigby2005, stasinopoulos2008, chavez2016, stasinopoulos2017}. We have then performed a series of hypothesis tests based on scoring rules, to evaluate the forecasting performance of severity models based on different cyber risk classifications \citep{gneiting2007A,gneiting2007B,rizzo2009, gneiting2011}. 

The findings of our study have several important implications for both cyber risk management and cyber risk insurance. While cyber risk classifications show promise in frequency modeling, their performance in severity modeling appears to be limited. This suggests that cyber insurance ratemakers could disregard the usage of cyber risk classifications in severity modeling, and then subsequently adjust insurance premiums based on frequency modeling dependent on the chosen classification. Moreover, one should exercise caution with projecting in-sample results for specific cyber risk types to out of sample applications. Second, our analysis highlights the challenges associated with using static classifications in a dynamic and rapidly evolving cyber risk landscape. Static cyber risk classifications, such as the operational risk based one, are found to be not flexible enough to capture the dynamic heterogeneity of cyber events. Moreover, despite efforts to construct dynamic classifications, such as the case of  impact-based approaches, the results indicate that these methods may still fall short in capturing the complex ever-changing nature of cyber risk adequately. Finally, while cyber risk classifications offer some benefits, particularly in frequency modeling, their effectiveness in severity modeling remains uncertain, especially in a forecast-oriented setting. However, our findings on the power of the conducted tests provide some indication that once larger datasets on losses from cyber events will be available, more revealing results on the adequate classification of risk might be achievable.   

To the best our knowledge, our study is the first to focus on distributional forecasts for cyber events. Future research in this area should focus on exploring more appropriate classification methods, without compromising the transparency and interpretability of cyber risk classifications. We conclude that correctly classifying, modeling and predicting the severity distribution of cyber  events remains an important and challenging issues for risk management and insurance.

\section*{Acknowledgement}
\noindent This research has been conducted within the Optus Macquarie University Cyber Security Hub and has been funded by its Risk Management, Governance and Control Program. The authors would like to acknowledge the participants of the American Risk and Insurance Association (ARIA) Annual Meeting 2024 for the valuable comments and feedback.

\section*{Data and Code}
\noindent Data were purchased from Advisen Ltd under the license agreement cannot be made available. The code to reproduce the results presented in the paper can be made available upon request. 
\clearpage
\appendix
\section{Mapping of Advisen cyber loss data between different classifications}
\label{appendix:mapping_classifications}
Table \ref{table:categorie1} shows the mapping between Advisen classification and the classification suggested by Eling and Romanosky.

   \begin{table}[h]
    	\centering
    	\caption[Categories]{This table reports the mapping from Advisen onto Eling or Romanosky classifications.}
    	\label{table:categorie1}
    
    	\begin{adjustbox}{width=\columnwidth,center}   
        	\begin{tabular}{lccccc}
        		Romanosky Classification                         & Data Breach &	Security Incident &	Privacy Violation &	Phishing Skimming &	Other\\
        		\hline

        		Privacy - Unauthorized Contact 	          &            &                     &         X           &                  &           \\
        		Privacy - Unauthorized Data 	          &            &                     &         X           &                  &           \\
        		Data - Physically Lost or Stolen          &     X      &                     &                     &                  &           \\
        		Identity - Fraudulent Use		          &     X      &                     &                     &                  &           \\
        		Data - Malicious Breach			          &     X      &                     &                     &                  &           \\
        		Phishing and Spoofing 			          &            &                     &                     &        X         &           \\
        		IT - Configuration/Implementation Errors  &            &          X          &                     &                  &           \\
        		Data - Unintentional Disclosure           &     X      &                     &                     &                  &           \\
        		Cyber Extortion					          &            &          X          &                     &                  &           \\
        		Network/Website Disruption		          &            &          X          &                     &                  &           \\
        		Skimming and Physical Tampering	          &            &                     &                     &        X         &           \\
        		IT - Processing Errors			          &            &          X          &                     &                  &           \\
        		Industrial Controls and Operation	      &            &          X          &                     &                  &           \\
        		Undetermined/Other				          &            &                     &                     &                  &     X     \\
        	 	\hline	

        		Eling Classification                                 & Actions by People &	System and Technical Failure &	Failed Internal Process &	External Event & Other\\
                \hline
        		Privacy - Unauthorized Contact 	          &                   &           X                  &                         &            &\\
        		Privacy - Unauthorized Data 	          &                   &           X                  &                         &            &\\
        		Data - Physically Lost or Stolen          &        X          &                              &                         &            &\\
        		Identity - Fraudulent Use		          &                   &           X                  &                         &            &\\
        		Data - Malicious Breach			          &        X           &                             &                         &            &\\
        		Phishing and Spoofing 			          &        X          &                              &                         &            &\\
        		IT - Configuration/Implementation Errors  &                   &                              &           X             &            &\\
        		Data - Unintentional Disclosure           &        X          &                              &                         &            &\\
        		Cyber Extortion					          &        X          &                              &                         &            &\\
        		Network/Website Disruption		          &                   &                              &           X             &            &\\
        		Skimming and Physical Tampering	          &        X          &                              &                         &            &\\
        		IT - Processing Errors			          &                   &                              &           X             &            &\\
        		Industrial Controls and Operation	      &                   &            X                 &                         &            &\\
        		Undetermined/Other				          &                   &                              &                         &            & X\\
        	 	\hline	
        	\end{tabular}
    	\end{adjustbox}
    \end{table}   

Advisen risk types are mapped onto the Frequency \& Severity risk types according to the their relative frequency and severity observed in the sample. First, Advisen risk types are ranked according to the their total number of events and median loss. Then, each risk type is classified as Rare if its number of events is between 0\% and 33\% (excluded) of the total numbers of events per risk type, Unlikely if its total number of events is between 33\% and 66\% (excluded) of the total numbers of events per risk type, and Likely if its total number of events is greater or equal than the 66\% of the total numbers of events per risk type. Events are associated with Low Severity, Medium Severity, and High Severity in a similar way in terms of median loss per risk type.

Table \ref{table:categorie2} shows the mapping between Advisen classification and the event types used in the Type \& Importance classification \citep[see,][]{acsc2020}. Events are allocated to risk types via a matrix-based classifier where event classes are also associate with the importance of the business sector the company affected operates. The three levels of importance are defined according to the median loss in each business sector: Low Importance corresponds to business sectors with median loss between 0\% and 33\% (excluded) of the sample, Medium Importance corresponds to business sectors with median loss between 33\% and 66\% (excluded) of the sample, and High Importance corresponds to business sectors with median losses greater or equal to the 66\% of the sample \citep[see,][]{shevchenko2023}. 

\begin{table}[h]
    	\centering
    	\caption[Categories]{This table reports the mapping from Advisen onto Type \& Importance event classes.}
    	\label{table:categorie2}
    
        	\begin{tabular}{lccc}
        		           & Low Level &	Exfiltration &	Disruption \\
        		\hline

        		Privacy - Unauthorized Contact 	          &    X       &              &         \\
        		Privacy - Unauthorized Data 	          &    X       &              &         \\
        		Data - Physically Lost or Stolen          &            &      X       &         \\
        		Identity - Fraudulent Use		          &            &      X       &         \\
        		Data - Malicious Breach			          &            &      X       &         \\
        		Phishing and Spoofing 			          &    X       &              &         \\
        		IT - Configuration/Implementation Errors  &            &              &    X    \\
        		Data - Unintentional Disclosure           &    X       &              &         \\
        		Cyber Extortion					          &            &      X       &         \\
        		Network/Website Disruption		          &            &              &    X    \\
        		Skimming and Physical Tampering	          &            &      X       &         \\
        		IT - Processing Errors			          &            &              &    X    \\
        		Industrial Controls and Operation	      &            &              &    X    \\
        		Undetermined/Other				          &            &              &         \\
        	 
        	 	\hline	
        	\end{tabular}
 \end{table}

\section{In-Sample Results}
Table \ref{tab:crps_testStats_insample} shows the values of the rCRPS based test statistic for the in-sample period. The null hypothesis is rejected, for all the weighting schemes in the case of the Advisen classification, consistently with the goodness-of-fit results reported in the Appendix \ref{appendix:goodness_of_fit}.
\begin{table}[h]
    \caption{This table shows the values of the test statistic based on rCRPS for the in-sample period. The critical values are 1.64 (5\% level of significance) and  2.32 (1\% level of significance).}
    \label{tab:crps_testStats_insample}
    \centering
    \begin{tabular}{l|rrrr}
    \toprule
    \multicolumn{5}{c}{Classification vs None Classification }\\
Classifications	&Equal&	Center&	Left	&Right\\
\midrule
Advisen	&1.88	&1.75	&1.79	&2.08\\
Romanosky	&-0.23&	-1.87&	-1.21&	0.28\\
Eling	&1.59	&1.50&	1.49	&1.48\\
Tail	&1.44&	1.44	&1.44	&1.43\\

Body	&1.46&	1.50	&1.48	&1.44\\
Frequency \& Severity	&1.42&	1.42&	1.42&	1.42\\
Type \& Importance	&-2.79	&1.70	&2.47&	-1.91\\
\midrule
    \multicolumn{5}{c}{Classification vs Random Classification }\\
Classifications	&Equal&	Center&	Left	&Right\\
\midrule
Advisen	&1.78	&1.80&	1.87&	1.92\\
Romanosky&	-0.44&	-2.12&-1.48&	0.55\\
Eling	&1.58	&1.52&	1.52&	1.48\\
Tail	&1.45&	1.45	&1.46&	1.44\\

Body	&1.46	&1.53&	1.51	&1.44\\
Frequency \& Severity	&1.42&	1.42	&1.42&	1.42\\
Type \& Importance	&1.58&	1.75&	3.25&	-2.15\\
\bottomrule
    \end{tabular}
\end{table}
Table \ref{tab:proportion_insample} shows the proportion of times the null hypothesis is rejected over the in-sample period. Results are comparable with the out of sample ones in Table \ref{tab:proportion}.
\begin{table}[h]
    \caption{Proportion of times, over the in-sample period, the null is rejected for the rCRPS test at the 5\% significance level.}
    \label{tab:proportion_insample}
    \centering
    \begin{tabular}{l|rrrr}
        \toprule

    \multicolumn{5}{c}{Classification vs None Classification }\\
    Classifications	&Equal&	Center&	Left	&Right\\
    \midrule
Advisen &0.11& 0.44 & 0.44 & 0.11\\
Romanosky &0.11 & 0.11 & 0.11 & 0.00\\
Eling & 0.22& 0.22& 0.11 & 0.33\\
Tail &0.11 & 0.11 & 0.11 & 0.00\\

Body & 0.00 & 0.11 & 0.11 & 0.00\\
Frequency \&Severity & 0.11 & 0.11 & 0.00 & 0.11\\
Type \& Importance &0.22 & 0.11 & 0.44 & 0.11\\
    \midrule
    \multicolumn{5}{c}{Classification vs Random Classification }\\
    Classifications	&Equal&	Center&	Left	&Right\\
    \midrule
Advisen & 0.11 & 0.33 & 0.22 & 0.11\\
Romanosky & 0.22 & 0.22 & 0.33 & 0.33\\
Eling & 0.33 & 0.11 & 0.00  & 0.00\\
Tail & 0.22 & 0.11 & 0.22 & 0.44\\

Body & 0.22 & 0.11 & 0.22 & 0.11\\
Frequency \& Severity & 0.11 & 0.00 &0.11 & 0.11\\
Type \& Importance & 0.22 & 0.11 & 0.22 & 0.00\\
\midrule    
    \end{tabular}
\end{table}

Similar in-sample results are reported in Tables \ref{tab:rES_testStats_insample} and \ref{tab:proportion_es_insample} for the case of the test statistic based on rES.

\begin{table}[h]
    \caption{This table shows the values of the test statistic based on rES for the in-sample period. The critical values are 1.64 (5\% level of significance) and  2.32 (1\% level of significance).}
    \label{tab:rES_testStats_insample}
    \centering
    \begin{tabular}{l|rrrr}
    \toprule
    \multicolumn{5}{c}{Classification vs None Classification }\\
Classifications	&Equal&	Center&	Left	&Right\\
\midrule
Advisen	&1.51	&1.49	&1.49	&1.49\\
Romanosky	&1.57	&1.79	&1.64	&1.48\\
Eling	&1.48	&1.46	&1.46	&1.45\\
Tail	&1.44	&1.43	&1.43	&1.44 \\

Body	&1.44	&1.44	&1.44	&1.43\\
Frequency \& Severity	&1.42	&1.42	&1.42	&1.42\\
Type \& Importance	&1.82	&1.89	&1.96	&8.39 \\
\midrule
    \multicolumn{5}{c}{Classification vs Random Classification }\\
Classifications	&Equal&	Center&	Left	&Right\\
\midrule

Advisen		&1.49		&1.49		&1.49	&	1.47\\
Romanosky 	&1.53		&1.87		&1.67		&1.47\\
Eling		&1.47	&	1.47		&1.46		&1.45\\
Tail		&1.45		&1.45		&1.45		&1.44\\

Body		&1.44		&1.45	&1.44		&1.43\\
Frequency \& Severity		&1.42		&1.42		&1.42		&1.42\\
Type \& Importance		&8.55		&1.90		&1.97	&2.56\\
\bottomrule
    \end{tabular}
\end{table}

\begin{table}[h]
    \caption{Proportion of times, over the in-sample period, the null is rejected for the rES test at the 5\% significance level.}
    \label{tab:proportion_es_insample}
    \centering
    \begin{tabular}{l|cccc}
    \toprule
    \multicolumn{5}{c}{Classification vs None Classification }\\
Classifications	&Equal&	Center&	Left	&Right\\
\midrule
Advisen& 0.44 &0.55 &0.55 &0.55\\
Romanosky &0.55 &0.11 &0.22 &0.55\\
Eling &0.22 &0.22 &0.33 &0.33\\

Tail &0.22 &0.22 &0.33 &0.22\\

Body &0.11&0.44 &0.44 &0.11\\
Frequency \& Severity &0.44 &0.33 &0.33 &0.44\\
Type \& Importance &0.33 &0.22 &0.22 &0.33\\
\midrule
    \multicolumn{5}{c}{Classification vs random Classification }\\
Classifications	&Equal&	Center&	Left	&Right\\
Advisen &0.55 &0.55 &0.55 &0.55\\
Romanosky &0.44 &0.33 &0.22&0.44\\
Eling &0.33 &0.33 &0.44 &0.33\\
Tail &0.22 &0.33 &0.44 &0.22\\

Body &0.11 &0.55 &0.55 &0.11\\
Frequency \& Severity &0.55 &0.33 &0.33 &0.44\\
Type \& Importance &0.33 &0.11 &0.22 &0.33\\
\midrule
    \end{tabular}
\end{table}

\begin{figure}[h]
    \centering
    \includegraphics[width = 0.9\textwidth, height = 0.4\columnwidth]{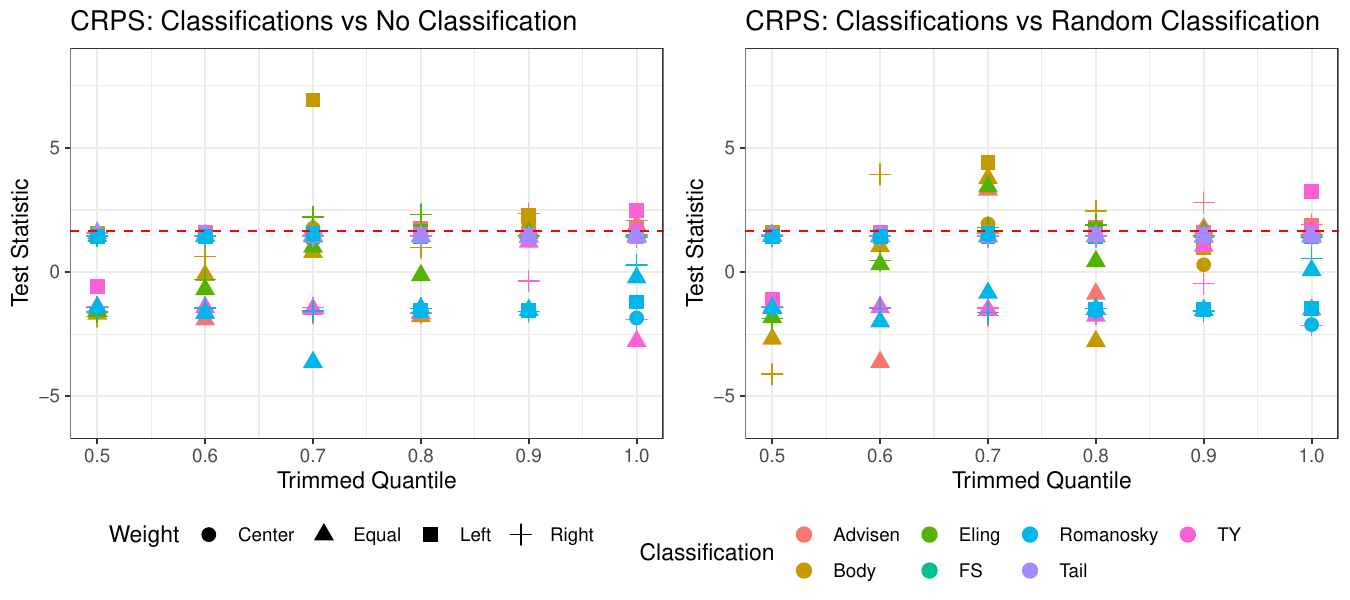}
    \caption{This figure shows the test statistics corresponding to rCRPS for different classifications, under different weighting schemes in the in-sample period. The trimmed quantiles are 50\%, 60\%, 70\%, 80\%,and 90\%. For completeness, the test statistic corresponding to the full sample is included as well. The red dashed line corresponds to the critical values for the one sided test at the 5\% level of significance.}
    \label{fig:trimmed_crps_insample}
\end{figure}

\begin{figure}[h]
    \centering
    \includegraphics[width = 0.9\textwidth, height = 0.4\columnwidth]{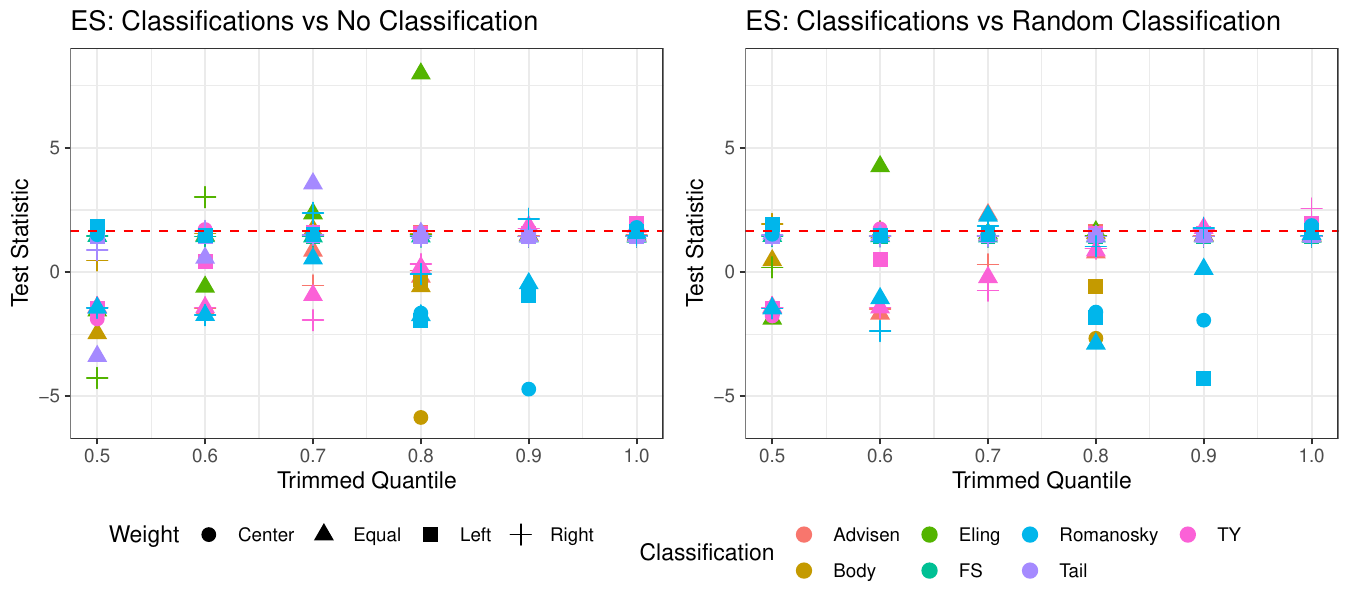}
    \caption{This figure shows the test statistics corresponding to rES for different classifications, under different weighting schemes in the in-sample period. The trimmed quantiles are 50\%, 60\%, 70\%, 80\%,and 90\%. For completeness, the test statistic corresponding to the full sample is included as well. The red dashed line corresponds to the critical values for the one sided test at the 5\% level of significance.}
    \label{fig:trimmed_es_insample}
\end{figure}

Figures \ref{fig:trimmed_crps_insample} and \ref{fig:trimmed_es_insample} show the results of the test of hypothesis on the trimmed in-sample observations. Consistently with the out of sample results, the trimming exercise in the in-sample period does not increase the out of sample performance of the model. 

\clearpage

\section{Goodness-of-Fit Tests}
\label{appendix:goodness_of_fit}
This appendix shows the results of goodness-of-fit tests for both the in-sample and the out of sample period. For every classification, we perform Kolmogorov-Smirnov (KS), Cramer-von Mises (CvM), and Anderson-Darling (AD) normality tests for the distribution of the standardised residuals\footnote{The standardised residuals are defined as $r_{i,t} = \Phi^{-1}\left(G(Y_{i,t},;\widehat{\mu}_{i,t|T},\widehat{\tau}_{i,t|T})\right)$ for the in-sample period and $r_{i,T+1|T} = \Phi^{-1}\left(G(Y_{i,T+1};\widehat{\mu}^1_{i,T+1|T},\widehat{\tau}_{i,T+1|T}^1)\right)$ in the out of sample period.}. Table \ref{tab:ks_cvm_ad_insample} shows the p-values of in the in-sample period, obtained via MonteCarlo simulation. Although threat based classifications seem to return a better performance during the in-sample period, as the weighting of the distance between distributions is shifted towards the tails, such as the cases of CvM and AD, the rejection rates increases. All the other classifications return a somewhat underwhelming results, with the Type \& Importance classification being the only one to exhibits consistent results across the three goodness-of-fit tests. A note of caution should be used in interpreting the results for the in-sample periods 2014/18, 2015/19, and 2016/20 since the high p-values reported in the table are likely due to small sample sizes and low statistical power.

Table \ref{tab:ks_cvm_ad_out} shows the p-values for the goodness-of-fit tests in the out of sample years. As expected, the null hypothesis is rejected more often than in the in-sample period. In this case, the impact based classifications are the only ones where the null hypothesis cannot be rejected during some of the out of sample periods.

\begin{table}[]
    \caption{This table shows the results for Kolmogorov-Smirnov, Cramer-von Mises, and Anderson-Darling normality tests for the in-sample standardised residuals. The number of simulations is 10,000, and $-$ indicates a p-value lower than 1\%.}
    \label{tab:ks_cvm_ad_insample}
    \centering
    \begin{adjustbox}{width=0.99\textwidth}
    \begin{tabular}{l|rrrrrrrrr}
    \toprule
    &\multicolumn{9}{c}{Kolmogorov-Smirnov}\\
    	 &2008/12	&2009/13	&2010/14	&2011/15	&2012/16	&2013/17	&2014/18	&2015/19	&2016/20\\
      \midrule
Advisen	                &0.0371	&0.0574	&0.0363	&0.0364	&0.0355	&0.0224	&0.1754	&0.6673	&0.3657\\
Romanosky	            &-	&-	&0.1268	&-	&0.0299	&0.1079	&0.1908	&0.4578	&0.2785\\
Eling	                &-	&-	&0.0105	&-&-	&0.0120	&0.0419	&0.1914	&0.1335\\
Tail	                &-	&-	&-	&-	&-	&0.0113	&0.1572	&0.0711	&0.1000\\
Body	                &-	&-	&0.0135	&-	&0.0152	&0.0253	&0.0582	&0.0292	&0.1186\\
Frequency \& Severity	&-	&-	&-	&-	&-	&0.1002	&-	&-	&0.1013\\
Type \& Importance	    &-	&-	&0.0524	&0.0323	&-	&-	&-	&0.3713	&0.8801\\
No Classification &-	&-	&-	&-	&-	&0.0564	&0.0871	&0.3105	&0.1665\\
Random Classification &-	&-	&- &-	&-	&0.0579	&0.1150	&0.1326	&0.0298\\
\midrule
&\multicolumn{9}{c}{Cramer-von Mises}\\
    	 &2008/12	&2009/13	&2010/14	&2011/15	&2012/16	&2013/17	&2014/18	&2015/19	&2016/20\\
\midrule
Advisen	&0.0129	&0.0551&	0.0105	&0.0161	&-	&-	&0.0255	&0.7002	&0.2738\\
Romanosky	&-	&-	&-	&-	&-	&-	&0.0768	&0.1337	&0.1028\\
Eling	&-	&-	&- &-	&-	&-	&0.029	&0.189	&0.0936\\
Tail	&-	&0.0184	&0.0389	&0.0145	&-	&0.0156	&0.0576	&0.4284	&0.2144\\
Body	&-	&-	&-	&-	&-	&0.0148	&0.0589	&0.1307	&0.1194\\
Frequency \& Severity	&-	&-	&-	&-	&-	&0.0233	&-	&-	&0.032\\
Type \& Importance	&-	&-	&0.1593	&0.0385	&-	&-	&-	&0.3461	&0.5293\\
No Classification &-	&-	&-	&-	&-	&0.0101&	0.0833	&0.1685	&0.098\\
Random Classification &-	&-	&0.0056	&0.0178	&-	&0.0153	&0.0638	&0.0479	&-\\
\midrule

&\multicolumn{9}{c}{Anderson-Darling}\\
    	 &2008/12	&2009/13	&2010/14	&2011/15	&2012/16	&2013/17	&2014/18	&2015/19	&2016/20\\
\midrule
Advisen	&-	&0.0853	&-	&-	&-	&-	&0.0276	&0.6245	&0.2498\\
Romanosky	&-	&0.0252	&0.0117	&-	&-	&-	&0.0413	&0.3084	&0.2095\\
Eling	&-	&-	&-	&- &-	&-	&0.0148	&0.0773	&0.0941\\
Tail	&-	&-	&-	&-	&-	&-	&0.0439	&0.0702	&0.1114\\

Body	&-	&-	&-	&-	&-	&-	&0.0251&	0.0541	&0.1383\\
Frequency \& Severity	&-	&-	&-	&-	&-	&-	&-	&-	&-\\
Type \& Importance	&-	&-	&0.0319	&0.0162	&-	&-	&-	&0.0964	&0.3689\\
No Classification     &-	&-	&-	&-	&	-&-	&0.0838	&0.1649	&0.1211\\ 
Random Classification &-	&-	&-	&-	&-	&-	&0.0708	&0.0721	&-\\
      \midrule
    \end{tabular}
    \end{adjustbox}
\end{table}

\begin{table}[]
    \caption{This table shows the results for Kolmogorov-Smirnov, Cramer-von Mises, and Anderson-Darling normality tests for the out of sample standardised residuals. The number of simulations is 10,000, and $-$ indicates a p-value lower than 1\%.}
    \label{tab:ks_cvm_ad_out}
    \centering
    \begin{adjustbox}{width=0.99\textwidth}
    \begin{tabular}{l|rrrrrrrrr}
    \toprule

    &\multicolumn{9}{c}{Kolmogorov-Smirnov}\\
    	 &2013	&2014	&2015	&2016	&2017	&2018	&2019	&2020	&2021\\
      \midrule
Advisen	                &-	&-	&-	&-	&-	&-	&-	&-	&0.0226\\
Romanosky	             &-	&-	&-	&-	&-	&-	&-	&-	&0.0243\\
Eling	                &-	&-	&-	&-	&-	&-	&-	&-	&0.155\\

Tail	                &-	&-	&-	&-	&-	&-	&-	&0.0594	&0.1093\\

Body	               &-	&-	&-	&-	&-	&-	&-	&-	&0.0545\\
Frequency \& Severity	&-	&-	&0.0168	&0.1046	&-	&0.7862	&0.1165	&0.048	&0.9139\\
Type \& Importance	    &-	&-	&0.9311	&0.2151	&-	&-	&-	&0.2953	&0.6319\\
No Classification    &-	&-	&-	&0.0109	&-	&-	&-	&-	&0.1222\\
Random Classification &-	&-	&-	&0.0161	&-	&-	&-	&-	&0.1583\\
\midrule
&\multicolumn{9}{c}{Cramer-von Mises}\\
    	 &2013	&2014	&2015	&2016	&2017	&2018	&2019	&2020	&2021\\
\midrule
Advisen	                &-	&-	&-	&-	&-	&-	&-	&-		&0.0208\\
Body	                &-	&-	&-	&-	&-	&-	&-	&-		&0.0202\\
Eling	                &-	&-	&-	&-	&-	&-	&-	&-		&0.0217\\
Frequency \& Severity	&-	&-	&0.0124	&0.0304	&-	&0.6694	&0.0105	&0.0136	&0.8617\\
Type \& Importance	    &-	&-	&0.9410	&0.4478	&-	&-	&-	&0.2054	&0.5358\\
Tail	                &-	&-	&-	&-	&-	&-	&-	&-	&	0.0906\\
Romanosky	            &-	&-	&-	&-	&-	&-	&-	&-		&0.0108\\
No Classification       &-	&-	&-	&-	&-	&-	&-	&-		&0.1580\\
Random Classification   &-	&-	&-	&-	&-	&-	&-	&-		&0.1556\\
\midrule
&\multicolumn{9}{c}{Anderson-Darling}\\
    	 &2013	&2014	&2015	&2016	&2017	&2018	&2019	&2020	&2021\\
\midrule
Advisen	                &-	&-	&-	&-	&-	&-	&-	&-	&-\\
Romanosky	             &-	&-	&-	&-	&-	&-	&-	&-	&- \\
Eling	                &-	&-	&-	&-	&-	&-	&-	&-	&-\\
Tail	                &-	&-	&-	&-	&-	&-	&-	&-	&0.0458\\
Body	               &-	&-	&-	&-	&-	&-	&-	&-	&-\\
Frequency \& Severity	&-	&-	&-	&-	&-	&0.409	&-	&-	&0.843\\
Type \& Importance	    &-	&-	&0.7909	&0.3419&	-	&-	&-	&0.1614	&0.5674\\
No Classification   	&-	&-	&-	&-	&-	&-	&-	&-	&0.1125\\
Random Classification 	&-	&-	&-	&-	&-	&-	&-	&-	&0.0545\\
      \midrule
    \end{tabular}
    \end{adjustbox}
\end{table}

\clearpage

\section{Out of Sample Performance Test with Lognormal Based Residual Scoring Functions.}
\label{appendix:rcrps_es_lognormal}
This appendix shows the results of a robustness analysis where the scoring functions in Equation (\ref{eq:res_scoring}) are based on a lognormal transformation (henceforth modified residual scoring rules):
\begin{align}
    &rCRPS(F,y) = \mathbb{E}\left[|W - F_{lnorm}^{-1}(F(y))|\right] -\dfrac{1}{2}\mathbb{E}\left[|W - W'|\right],\notag\\
    &rES(F,y) =\mathbb{E}\left[|W - F_{lnorm}^{-1}(F(y))|^{\beta}\right] - \dfrac{1}{2}\mathbb{E}\left[|W - W'|^{\beta}\right],\notag
\end{align}
where $W$ and $W'$ are independent copies of a random variable distributed as a lognormal, with mean equal to 1 and standard deviation equal to $e$, and $F_{lnorm}^{-1}$ is the inverse of the cumulative distribution function of lognormal random variable with the same set of parameters. One should note that the statistical power of the test hypothesis in Equation (\ref{eq:test}) strongly depends on the standard deviation of the test statistic in Equation (\ref{eq:test_statistic}). Therefore it is expected for the test based on the modified residual scoring rules to exhibit lower level of power than the non-modified counterparts.

\begin{figure}[H]
    \centering
    \includegraphics[width = 0.9\textwidth, height = 0.7\columnwidth]{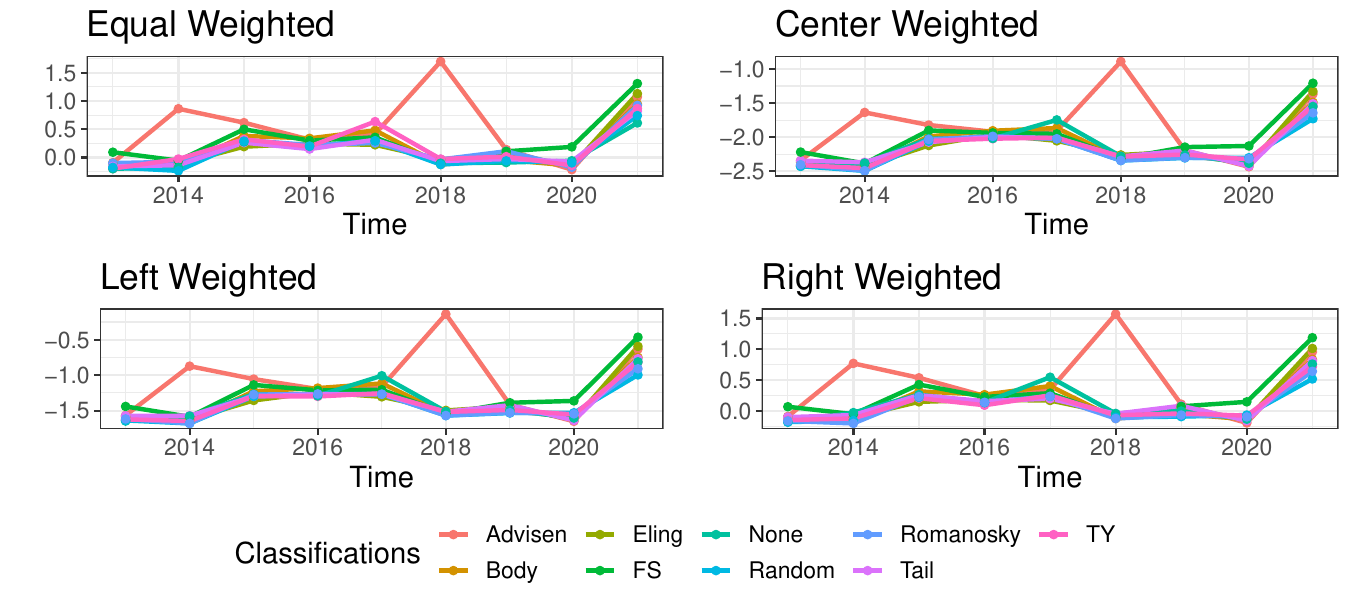}
    \caption{This figure shows the yearly averages of rCRPS based on lognormal transformation for different classification for 4 different weighting functions on the log-scale. Random and None classifications are included for comparison. Adivsen classification and Type \& Importance show a better forecasting performance than other classifications. FS and TY refer to Frequency \& Severity and Type \& Importance classifications, respectively.}
    \label{fig:crps_outsample_appendix}
\end{figure}

\begin{figure}[H]
    \centering
    \includegraphics[width = 0.9\textwidth, height = 0.7\columnwidth]{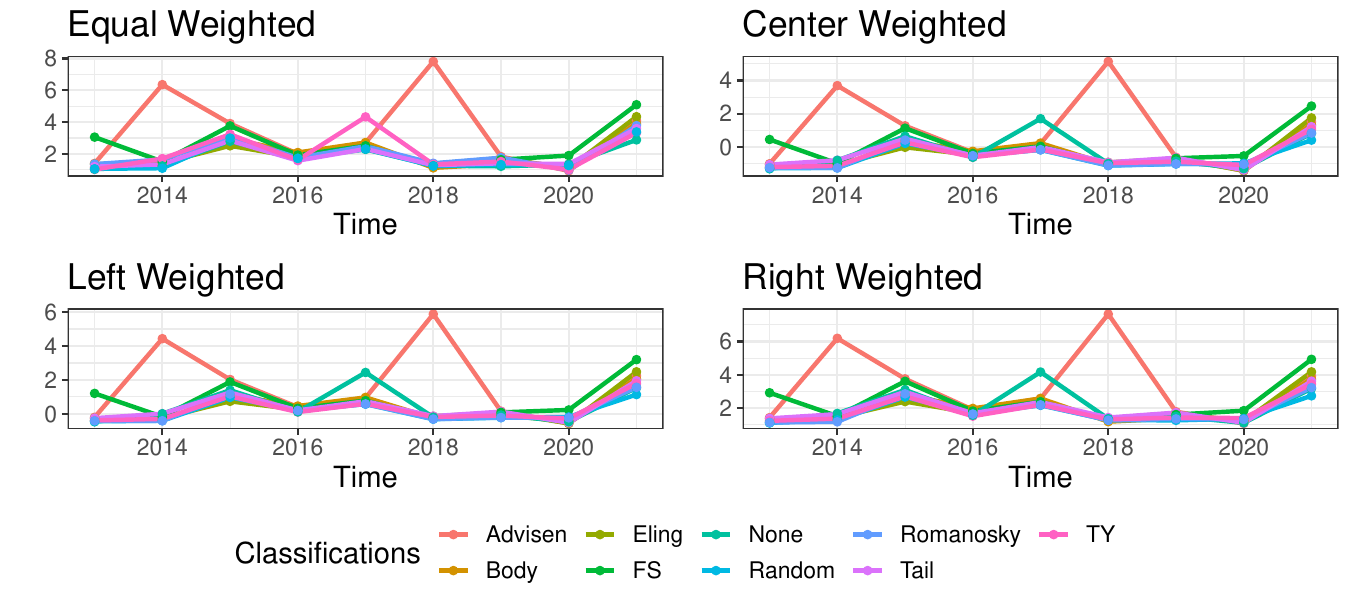}
    \caption{This figure shows the yearly averages of rES based on lognormal transformation for different classification for 4 different weighting functions on the log-scale. Random and None classifications are included for comparison. Adivsen classification and Type \& Importance show a better forecasting performance than other classifications. FS and TY refer to Frequency \& Severity and Type \& Importance classifications, respectively.}
    \label{fig:es_outsample_appendix}
\end{figure}

Figures \ref{fig:crps_outsample_appendix} and \ref{fig:es_outsample_appendix} depict the yearly average values of the modified residual scoring rules for different classifications, according to different weighting functions. The overall behavior of the yearly average values is similar to the one of the residual scoring rules, with the Advisen classification returning higher values in 2014 and 2018 in both the modified rCRPS and rES. As expected the values for both the modified rCRPS and rES range in wider intervals than the values depicted in Figures \ref{fig:crps_outsample} and \ref{fig:es_outsample}.

Similarly to the tests based on residual scoring rules, Tables \ref{tab:crps_ts_outsample_m} and \ref{tab:crps_ts_outsample_m2} report the values of the test statistics based on the modified rCRPS and rES, respectively. According to the results, when the test statistic is based on the modified residual scoring rules, the null hypothesis gets rejected more often, and in fact it appears that it would be possible to distinguish the forecasting performance of models based on cyber risk classifications from a random classification  based model and from a model without any classification. However this result is heavily affected by the limited sample sizes and the artificially increased standard deviation of the test statistic based on the lognormal transformation. To verify this, Tables \ref{tab:power_rES_m} and \ref{tab:power_rES_m2} report the results of a power analysis in a similar set up to the Subsection \ref{sec:power}. As expected, the power of both tests is considerably lower than one based on the residual scoring rules, especially in the case of the right weighted scoring rules, which are arguably the most relevant cases for risk management and insurance purposes. We also have performed a similar robustness analysis using a skew normal distribution with parameters $SN(0,1,5)$ in the modified scoring rules. The results are similar to the ones corresponding to the residual scoring rules, confirming that the apparent superior performance reported in Table \ref{tab:crps_ts_outsample_m} and \ref{tab:crps_ts_outsample_m2} is due to the sensitivity of the lognormal distribution to heavy tailed data rather than to features of the cyber risk classifications.

\begin{table}[H]
    \caption{This table shows the values of the test statistic based on modified rCRPS for the overall out of sample period. The critical values are 1.64 (5\% level of significance) and  2.32 (1\% level of significance).}
    \label{tab:crps_ts_outsample_m}
    \centering
    \begin{tabular}{l|rrrr}
    \toprule
    \multicolumn{5}{c}{Classification vs None Classification }\\
Classifications	&Equal&	Center&	Left	&Right\\
\midrule
Advisen	&1.71	&1.69	&1.69	&1.71\\
Romanosky	&1.51	&1.49	&1.49	&1.50\\ 
Eling	& 1.57	&1.55	&1.56	&1.57\\
Tail	&1.60	&1.51	&1.53	&1.61\\

Body	& 1.46	&1.45	&1.45	&1.46\\
Frequency \& Severity	& 1.52	&1.50	&1.51	&1.52\\
Type \& Importance	&  1.79	&1.81	&1.86	&1.78\\
\midrule
\multicolumn{5}{c}{Classification vs Random Classification }\\
Classifications	&Equal&	Center&	Left	&Right\\
\midrule

Advisen &	1.71	&1.71	&1.71	&1.71\\
Romanosky	& 1.51	&1.54	&1.55	&1.50\\
Eling	& 1.58	&1.68	&1.73	&1.58\\

Tail	& 1.60	&1.89	&2.21	&1.61\\

Body	& 1.46	&1.47	&1.47	&1.46\\
Frequency \& Severity	& 1.52	&1.51	&1.52	&1.52\\
Type \& Importance	& 1.79	&2.04	&2.15	&1.78\\
\bottomrule
\end{tabular}
\end{table}

\begin{table}[H]
  \caption{This table shows the values of the test statistic based on modified rES for the overall out of sample period. The critical values are 1.64 (5\% level of significance) and  2.32 (1\% level of significance).}
    \label{tab:crps_ts_outsample_m2}
    \centering
    \begin{tabular}{l|rrrr}
    \toprule
    \multicolumn{5}{c}{Classification vs None Classification }\\
Classifications	&Equal&	Center&	Left	&Right\\
\midrule
Advisen	& 2.03	&2.02	&2.02	&2.03\\
Romanosky	& 1.53	&1.53	&1.52	&1.52\\ 
Eling	& 2.07	&2.02	&2.06	&2.07\\
Tail	& 1.83	&1.71	&1.73	&1.80\\

Body	& 1.51	&1.49	&1.49	&1.51\\
Frequency \& Severity	& 1.69&	1.67	&1.68	&1.69\\
Type \& Importance	& 2.29&	2.25	&2.25	&2.28 \\
\midrule
\multicolumn{5}{c}{Classification vs Random Classification }\\
Classifications	&Equal&	Center&	Left	&Right\\
\midrule

Advisen &	2.03	&2.03	&2.03	&2.03\\
Romanosky	& 1.57	&1.58	&1.58	&1.56\\
Eling	& 1.54	&1.37	&1.27	&1.52\\
Tail	& 0.24	&0.19	&0.15	&0.26\\

Body	& 1.82	&1.73	&1.75	&1.79\\
Frequency \& Severity	&1.73	&1.72	&1.73	&1.73 \\
Type \& Importance	& 2.59	&2.59	&2.59	&2.56 \\
\bottomrule
\end{tabular}
\end{table}

\begin{table}[H]
    \caption{This table shows the results of the power analysis in the case of modified rCRPS. For each sample size, 10,000 random draws are sampled from the generated losses due cyber events. The power of the test is computed as the fraction of random draws for which the null hypothesis of equal performance is rejected.}
    \label{tab:power_rES_m}
    \centering
    \begin{tabular}{l|rrrr}
    \toprule
    \multicolumn{5}{c}{Type \& Importance vs None Classification }\\
        Sample Size & Equal& Left& Center & Right \\ 
            \midrule
20  &0	&0.844	&0.568	&0\\
50	&0	&0.96	&0.9	&0\\
100	&0	&0.994	&0.982	&0\\
500	&0	&1	&1	&0\\
1000	&0	&1	&1	&0\\
2000	&0	&1	&1	&0\\
5000	&  0&1	&1	&0 \\
\midrule
\multicolumn{5}{c}{Type \& Importance vs Random Classification }\\
        Sample Size & Equal& Left& Center & Right \\ 
\midrule
20 	&0	&0.836	&0.566	&0\\
50 	&0	&0.96	&0.894	&0\\
100  	&0	&0.994	&0.974	&0\\
500	&0	&1	&1	&0\\
1000	&0	&1	&1	&0\\
2000	&0	&1	&1	&0\\
5000	&  0&1	&1	&0 \\
\bottomrule
    \end{tabular}
\end{table}

\begin{table}[H]
  \caption{This table shows the results of the power analysis in the case of modified rES. For each sample size, 10,000 random draws are sampled from the generated losses due cyber events. The power of the test is computed as the fraction of random draws for which the null hypothesis of equal performance is rejected.}
    \label{tab:power_rES_m2}
    \centering
    \begin{tabular}{l|rrrr}
    \toprule
    \multicolumn{5}{c}{Type \& Importance vs None Classification }\\
        Sample Size & Equal& Left& Center & Right \\ 
            \midrule
20  &0	&0.170	&0.052	&0\\
50	&0	&0.362	&0.132	&0\\
100	&0	&0.582	&0.310	&0\\
500	&0	&0.810	&0.612	&0\\
1000	&0	&0.836	&0.686	&0\\
2000	&0	&0.871	&0.764	&0\\
5000	&0	&0.920	&0.782	&0\\
\midrule
\multicolumn{5}{c}{Type \& Importance vs Random Classification }\\
        Sample Size & Equal& Left& Center & Right \\ 
\midrule
20 	&0	&0.170	&0.060	&0\\
50 	&0	&0.362	&0.128	&0\\
100  	&0	&0.566	&0.314	&0\\
500	&0	&0.788	&0.580	&0\\
1000	&0	&0.804	&0.656	&0\\
2000	&0	&0.842	&0.723	&0\\
5000	&0	&0.870	&0.783	&0\\
\bottomrule
    \end{tabular}
\end{table}
\clearpage
\bibliography{mybibfile}

\begin{thebibliography}{}

\bibitem[\protect\citeauthoryear{ACSC}{ACSC}{2021}]{acsc2020}
ACSC (2021).
\newblock Australian cyber security centre: Annual cyber threat report.

\bibitem[\protect\citeauthoryear{Adelmann, Ergen, Gaidosch, Jenkinson,
  Khiaonarong, Morozova, Schwarz, and Wilson}{Adelmann
  et~al.}{2020}]{adelmann2020cyber}
Adelmann, F., I.~Ergen, T.~Gaidosch, N.~Jenkinson, M.~T. Khiaonarong,
  A.~Morozova, N.~Schwarz, and C.~Wilson (2020).
\newblock {\em Cyber risk and financial stability: It’sa small world after
  all}.
\newblock International Monetary Fund.

\bibitem[\protect\citeauthoryear{Aldasoro, Gambacorta, Giudici, and
  Leach}{Aldasoro et~al.}{2020}]{aldasoro2020}
Aldasoro, I., L.~Gambacorta, P.~Giudici, and T.~Leach (2020).
\newblock The drivers of cyber risk.
\newblock Technical report, Bank of International Settlements.

\bibitem[\protect\citeauthoryear{Aldasoro, Gambacorta, Giudici, and
  Leach}{Aldasoro et~al.}{2022}]{aldasoro2022}
Aldasoro, I., L.~Gambacorta, P.~Giudici, and T.~Leach (2022).
\newblock The drivers of cyber risk.
\newblock {\em Journal of Financial Stability\/}~{\em 60}, 100989.

\bibitem[\protect\citeauthoryear{Alexander, Coulon, Han, and Meng}{Alexander
  et~al.}{2022}]{alexander2022}
Alexander, C., M.~Coulon, Y.~Han, and X.~Meng (2022).
\newblock Evaluating the discrimination ability of proper multi-variate scoring
  rules.
\newblock {\em Annals of Operations Research\/}, 1--27.

\bibitem[\protect\citeauthoryear{Amisano and Giacomini}{Amisano and
  Giacomini}{2007}]{amisano2007}
Amisano, G. and R.~Giacomini (2007).
\newblock Comparing density forecasts via weighted likelihood ratio tests.
\newblock {\em Journal of Business \& Economic Statistics\/}~{\em 25\/}(2),
  177--190.

\bibitem[\protect\citeauthoryear{Balkema and De~Haan}{Balkema and
  De~Haan}{1974}]{balkema1974}
Balkema, A.~A. and L.~De~Haan (1974).
\newblock Residual life time at great age.
\newblock {\em The Annals of Probability\/}, 792--804.

\bibitem[\protect\citeauthoryear{Biener, Eling, and Wirfs}{Biener
  et~al.}{2015}]{biener2015}
Biener, C., M.~Eling, and J.~H. Wirfs (2015).
\newblock Insurability of cyber risk: An empirical analysis.
\newblock {\em The Geneva Papers on Risk and Insurance-Issues and
  Practice\/}~{\em 40\/}(1), 131--158.

\bibitem[\protect\citeauthoryear{Braun}{Braun}{1980}]{braun1980}
Braun, H. (1980).
\newblock A simple method for testing goodness of fit in the presence of
  nuisance parameters.
\newblock {\em Journal of the Royal Statistical Society: Series B
  (Methodological)\/}~{\em 42\/}(1), 53--63.

\bibitem[\protect\citeauthoryear{Cebula, Popeck, and Young}{Cebula
  et~al.}{2014}]{cebula2014}
Cebula, J.~J., M.~E. Popeck, and L.~R. Young (2014).
\newblock A taxonomy of operational cyber security risks version 2.
\newblock Technical report, Carnegie-Mellon Univ Pittsburgh Pa Software
  Engineering Inst.

\bibitem[\protect\citeauthoryear{Cebula and Young}{Cebula and
  Young}{2010}]{cebula2010}
Cebula, J.~L. and L.~R. Young (2010).
\newblock A taxonomy of operational cyber security risks.
\newblock Technical report, Carnegie-Mellon University, Pittsburgh Pa Software
  Engineering Inst.

\bibitem[\protect\citeauthoryear{Chavez-Demoulin, Embrechts, and
  Hofert}{Chavez-Demoulin et~al.}{2016}]{chavez2016}
Chavez-Demoulin, V., P.~Embrechts, and M.~Hofert (2016).
\newblock An extreme value approach for modeling operational risk losses
  depending on covariates.
\newblock {\em Journal of Risk and Insurance\/}~{\em 83\/}(3), 735--776.

\bibitem[\protect\citeauthoryear{{Chief Risk Officers Forum}}{{Chief Risk
  Officers Forum}}{2017}]{cro2017cro}
{Chief Risk Officers Forum} (2017).
\newblock Cro forum concept paper on a proposed categorisation methodology for
  cyber risk.
\newblock CRO Forum Amsterdam.

\bibitem[\protect\citeauthoryear{Cohen, Humphries, Veau, and Francis}{Cohen
  et~al.}{2019}]{cohen2019}
Cohen, R.~D., J.~Humphries, S.~Veau, and R.~Francis (2019).
\newblock An investigation of cyber loss data and its links to operational
  risk.
\newblock {\em Journal of Operational Risk\/}~{\em 14\/}(3), 1--25.

\bibitem[\protect\citeauthoryear{Cope and Labbi}{Cope and
  Labbi}{2008}]{cope2008}
Cope, E. and A.~Labbi (2008).
\newblock Operational loss scaling by exposure indicators: evidence from the
  {ORX} database.
\newblock {\em Journal of Operational Risk\/}~{\em 3\/}(4), 25--46.

\bibitem[\protect\citeauthoryear{Cs{\"o}rg{\H{o}} and Faraway}{Cs{\"o}rg{\H{o}}
  and Faraway}{1996}]{csorgHo1996}
Cs{\"o}rg{\H{o}}, S. and J.~J. Faraway (1996).
\newblock The exact and asymptotic distributions of {C}ram{\'e}r-von {M}ises
  statistics.
\newblock {\em Journal of the Royal Statistical Society: Series B
  (Methodological)\/}~{\em 58\/}(1), 221--234.

\bibitem[\protect\citeauthoryear{Curti, Gerlach, Kazinnik, Lee, and
  Mihov}{Curti et~al.}{2023}]{curti2023cyber}
Curti, F., J.~Gerlach, S.~Kazinnik, M.~Lee, and A.~Mihov (2023).
\newblock Cyber risk definition and classification for financial risk
  management.
\newblock {\em Journal of Operational Risk\/}~{\em 18\/}(2), 37--58.

\bibitem[\protect\citeauthoryear{{Cyentia Institute}}{{Cyentia
  Institute}}{2020}]{cyentia2020}
{Cyentia Institute} (2020).
\newblock A clearer vision for assessing the risk of cyber incidents.
\newblock Technical report, Cyentia Institute.

\bibitem[\protect\citeauthoryear{Dacorogna, Debbabi, and Kratz}{Dacorogna
  et~al.}{2023}]{dacorogna2023building}
Dacorogna, M., N.~Debbabi, and M.~Kratz (2023).
\newblock Building up cyber resilience by better grasping cyber risk via a new
  algorithm for modelling heavy-tailed data.
\newblock {\em European Journal of Operational Research\/}.

\bibitem[\protect\citeauthoryear{Dacorogna and Kratz}{Dacorogna and
  Kratz}{2023}]{dacorogna2023managing}
Dacorogna, M. and M.~Kratz (2023).
\newblock Managing cyber risk, a science in the making.
\newblock {\em Scandinavian Actuarial Journal\/}, 1--22.

\bibitem[\protect\citeauthoryear{Dahen and Dionne}{Dahen and
  Dionne}{2010}]{dahen2010}
Dahen, H. and G.~Dionne (2010).
\newblock Scaling models for the severity and frequency of external operational
  loss data.
\newblock {\em Journal of Banking \& Finance\/}~{\em 34\/}(7), 1484--1496.

\bibitem[\protect\citeauthoryear{Duffie and Younger}{Duffie and
  Younger}{2019}]{duffie2019cyber}
Duffie, D. and J.~Younger (2019).
\newblock Cyber runs: How a cyber attack could affect us financial
  institutions.

\bibitem[\protect\citeauthoryear{Edwards, Hofmeyr, and Forrest}{Edwards
  et~al.}{2016}]{edwards2016}
Edwards, B., S.~Hofmeyr, and S.~Forrest (2016).
\newblock Hype and heavy tails: A closer look at data breaches.
\newblock {\em Journal of Cybersecurity\/}~{\em 2\/}(1), 3--14.

\bibitem[\protect\citeauthoryear{Eisenbach, Kovner, and Lee}{Eisenbach
  et~al.}{2022}]{eisenbach2022cyber}
Eisenbach, T.~M., A.~Kovner, and M.~J. Lee (2022).
\newblock Cyber risk and the us financial system: A pre-mortem analysis.
\newblock {\em Journal of Financial Economics\/}~{\em 145\/}(3), 802--826.

\bibitem[\protect\citeauthoryear{Eling}{Eling}{2020}]{eling2020cyber}
Eling, M. (2020).
\newblock Cyber risk research in business and actuarial science.
\newblock {\em European Actuarial Journal\/}, 1--31.

\bibitem[\protect\citeauthoryear{Eling and Jung}{Eling and
  Jung}{2018}]{eling2018}
Eling, M. and K.~Jung (2018).
\newblock Copula approaches for modeling cross-sectional dependence of data
  breach losses.
\newblock {\em Insurance: Mathematics and Economics\/}~{\em 82}, 167--180.

\bibitem[\protect\citeauthoryear{Eling and Loperfido}{Eling and
  Loperfido}{2017}]{eling2017}
Eling, M. and N.~Loperfido (2017).
\newblock Data breaches: Goodness of fit, pricing, and risk measurement.
\newblock {\em Insurance: Mathematics and Economics\/}~{\em 75}, 126--136.

\bibitem[\protect\citeauthoryear{Eling, McShane, and Nguyen}{Eling
  et~al.}{2021}]{eling2021}
Eling, M., M.~McShane, and T.~Nguyen (2021).
\newblock Cyber risk management: History and future research directions.
\newblock {\em Risk Management and Insurance Review\/}~{\em 24\/}(1), 93--125.

\bibitem[\protect\citeauthoryear{Eling and Wirfs}{Eling and
  Wirfs}{2019}]{eling2019}
Eling, M. and J.~Wirfs (2019).
\newblock What are the actual costs of cyber risk events?
\newblock {\em European Journal of Operational Research\/}~{\em 272\/}(3),
  1109--1119.

\bibitem[\protect\citeauthoryear{Eling and Wirfs}{Eling and
  Wirfs}{2015}]{eling2015modelling}
Eling, M. and J.~H. Wirfs (2015).
\newblock Modelling and management of cyber risk.
\newblock {\em International Actuarial Association Life Section\/}.

\bibitem[\protect\citeauthoryear{Evans, Maglaras, He, and Janicke}{Evans
  et~al.}{2016}]{evans2016}
Evans, M., L.~A. Maglaras, Y.~He, and H.~Janicke (2016).
\newblock Human behaviour as an aspect of cybersecurity assurance.
\newblock {\em Security and Communication Networks\/}~{\em 9\/}(17),
  4667--4679.

\bibitem[\protect\citeauthoryear{Farkas, Lopez, and Thomas}{Farkas
  et~al.}{2021}]{farkas2021cyber}
Farkas, S., O.~Lopez, and M.~Thomas (2021).
\newblock Cyber claim analysis using {Generalized} {Pareto} regression trees
  with applications to insurance.
\newblock {\em Insurance: Mathematics and Economics\/}~{\em 98}, 92--105.

\bibitem[\protect\citeauthoryear{Ganegoda and Evans}{Ganegoda and
  Evans}{2013}]{ganegoda2013}
Ganegoda, A. and J.~Evans (2013).
\newblock A scaling model for severity of operational losses using generalized
  additive models for location scale and shape (gamlss).
\newblock {\em Annals of Actuarial Science\/}~{\em 7\/}(1), 61--100.

\bibitem[\protect\citeauthoryear{Gatzert and Schubert}{Gatzert and
  Schubert}{2022}]{gatzert2022}
Gatzert, N. and M.~Schubert (2022).
\newblock Cyber risk management in the us banking and insurance industry: A
  textual and empirical analysis of determinants and value.
\newblock {\em Journal of Risk and Insurance\/}~{\em 89\/}(3), 725--763.

\bibitem[\protect\citeauthoryear{Gneiting, Balabdaoui, and Raftery}{Gneiting
  et~al.}{2007}]{gneiting2007A}
Gneiting, T., F.~Balabdaoui, and A.~E. Raftery (2007).
\newblock Probabilistic forecasts, calibration and sharpness.
\newblock {\em Journal of the Royal Statistical Society Series B: Statistical
  Methodology\/}~{\em 69\/}(2), 243--268.

\bibitem[\protect\citeauthoryear{Gneiting and Raftery}{Gneiting and
  Raftery}{2007}]{gneiting2007B}
Gneiting, T. and A.~E. Raftery (2007).
\newblock Strictly proper scoring rules, prediction, and estimation.
\newblock {\em Journal of the American Statistical Association\/}~{\em
  102\/}(477), 359--378.

\bibitem[\protect\citeauthoryear{Gneiting and Ranjan}{Gneiting and
  Ranjan}{2011}]{gneiting2011}
Gneiting, T. and R.~Ranjan (2011).
\newblock Comparing density forecasts using threshold-and quantile-weighted
  scoring rules.
\newblock {\em Journal of Business \& Economic Statistics\/}~{\em 29\/}(3),
  411--422.

\bibitem[\protect\citeauthoryear{Healey, Mosser, Rosen, and Tache}{Healey
  et~al.}{2018}]{healey2018future}
Healey, J., P.~Mosser, K.~Rosen, and A.~Tache (2018).
\newblock The future of financial stability and cyber risk.
\newblock {\em The Brookings Institution Cybersecurity Project\/}, 1--18.

\bibitem[\protect\citeauthoryear{Jung}{Jung}{2021}]{jung2021}
Jung, K. (2021).
\newblock Extreme data breach losses: An alternative approach to estimating
  probable maximum loss for data breach risk.
\newblock {\em North American Actuarial Journal\/}, 1--24.

\bibitem[\protect\citeauthoryear{Kashyap and Wetherilt}{Kashyap and
  Wetherilt}{2019}]{kashyap2019some}
Kashyap, A.~K. and A.~Wetherilt (2019).
\newblock Some principles for regulating cyber risk.
\newblock In {\em AEA Papers and Proceedings}, Volume 109, pp.\  482--487.

\bibitem[\protect\citeauthoryear{Maillart and Sornette}{Maillart and
  Sornette}{2010}]{maillart2010}
Maillart, T. and D.~Sornette (2010).
\newblock Heavy-tailed distribution of cyber-risks.
\newblock {\em The European Physical Journal B\/}~{\em 75\/}(3), 357--364.

\bibitem[\protect\citeauthoryear{Malavasi, Peters, Shevchenko, Tr{\"u}ck, Jang,
  and Sofronov}{Malavasi et~al.}{2022}]{malavasi2022}
Malavasi, M., G.~W. Peters, P.~V. Shevchenko, S.~Tr{\"u}ck, J.~Jang, and
  G.~Sofronov (2022).
\newblock Cyber risk frequency, severity and insurance viability.
\newblock {\em Insurance: Mathematics and Economics\/}~{\em 106}, 90--114.

\bibitem[\protect\citeauthoryear{Nai-Fovino, Neisse, Hern{\'a}ndez-Ramos,
  Polemi, Ruzzante, Figwer, and Lazari}{Nai-Fovino et~al.}{2019}]{nai2019}
Nai-Fovino, I., R.~Neisse, J.~Hern{\'a}ndez-Ramos, N.~Polemi, G.~Ruzzante,
  M.~Figwer, and A.~Lazari (2019).
\newblock A proposal for a european cybersecurity taxonomy.
\newblock {\em Publications Office of the European Union\/}.

\bibitem[\protect\citeauthoryear{Nai-Fovino, Neisse, Lazari, Ruzzante, Polemi,
  and Figwer}{Nai-Fovino et~al.}{2018}]{nai2018}
Nai-Fovino, I., R.~Neisse, A.~Lazari, G.~Ruzzante, N.~Polemi, and M.~Figwer
  (2018).
\newblock European cybersecurity centres of expertise map - definitions and
  taxonomy.
\newblock {\em Publications Office of the European Union\/}.

\bibitem[\protect\citeauthoryear{NIST}{NIST}{2004}]{pub2004standards}
NIST (2004).
\newblock Standards for security categorization of federal information and
  information systems.
\newblock Federal Information Processing Standards Publication, National
  Institute of Standards and Technology, FIPS PUB 199.

\bibitem[\protect\citeauthoryear{Peters, Malavasi, Sofronov, Shevchenko,
  Tr{\"u}ck, and Jang}{Peters et~al.}{2023}]{peters2023}
Peters, G.~W., M.~Malavasi, G.~Sofronov, P.~V. Shevchenko, S.~Tr{\"u}ck, and
  J.~Jang (2023).
\newblock Cyber loss model risk translates to premium mispricing and risk
  sensitivity.
\newblock {\em The Geneva Papers on Risk and Insurance-Issues and
  Practice\/}~{\em 48\/}(2), 372--433.

\bibitem[\protect\citeauthoryear{Peters, Shevchenko, and Cohen}{Peters
  et~al.}{2018}]{peters2018}
Peters, G.~W., P.~V. Shevchenko, and R.~D. Cohen (2018).
\newblock Understanding cyber-risk and cyber-insurance.
\newblock In D.~Maurice, D.~Fairman, and J.~Freund (Eds.), {\em Fintech: Growth
  and Deregulation}, Chapter~12, pp.\  303--330. United Kingdom: Risk Books.

\bibitem[\protect\citeauthoryear{Peters, Shevchenko, Cohen, and Maurice}{Peters
  et~al.}{2018}]{peters2017}
Peters, G.~W., P.~V. Shevchenko, R.~D. Cohen, and D.~R. Maurice (2018).
\newblock Statistical machine learning analysis of cyber risk data: event case
  studies.
\newblock In D.~Maurice, D.~Fairman, and J.~Freund (Eds.), {\em Fintech: Growth
  and Deregulation}, Chapter~3, pp.\  75--99. United Kingdom: Risk Books.

\bibitem[\protect\citeauthoryear{Pickands}{Pickands}{1975}]{pickands1975}
Pickands, J. (1975).
\newblock Statistical inference using extreme order statistics.
\newblock {\em The Annals of Statistics\/}, 119--131.

\bibitem[\protect\citeauthoryear{Rigby and Stasinopoulos}{Rigby and
  Stasinopoulos}{2005}]{rigby2005}
Rigby, R. and D.~Stasinopoulos (2005).
\newblock Generalized additive models for location, scale and shape.
\newblock {\em Journal of the Royal Statistical Society: Series C (Applied
  Statistics)\/}, 507–554.

\bibitem[\protect\citeauthoryear{Rizzo}{Rizzo}{2009}]{rizzo2009}
Rizzo, M.~L. (2009).
\newblock New goodness-of-fit tests for {Pareto} distributions.
\newblock {\em ASTIN Bulletin: The Journal of the IAA\/}~{\em 39\/}(2),
  691--715.

\bibitem[\protect\citeauthoryear{Romanosky}{Romanosky}{2016}]{romanosky2016}
Romanosky, S. (2016).
\newblock Examining the costs and causes of cyber incidents.
\newblock {\em Journal of Cybersecurity\/}~{\em 2\/}(2), 121--135.

\bibitem[\protect\citeauthoryear{Romanosky, Ablon, Kuehn, and Jones}{Romanosky
  et~al.}{2019}]{romanosky2019content}
Romanosky, S., L.~Ablon, A.~Kuehn, and T.~Jones (2019).
\newblock Content analysis of cyber insurance policies: How do carriers price
  cyber risk?
\newblock {\em Journal of Cybersecurity\/}~{\em 5\/}(1), tyz002.

\bibitem[\protect\citeauthoryear{Selten}{Selten}{1998}]{selten1998}
Selten, R. (1998).
\newblock Axiomatic characterization of the quadratic scoring rule.
\newblock {\em Experimental Economics\/}~{\em 1}, 43--61.

\bibitem[\protect\citeauthoryear{Shetty, McShane, Zhang, Kesan, Kamhoua, Kwiat,
  and Njilla}{Shetty et~al.}{2018}]{shetty2018}
Shetty, S., M.~McShane, L.~Zhang, J.~P. Kesan, C.~A. Kamhoua, K.~Kwiat, and
  L.~L. Njilla (2018).
\newblock Reducing informational disadvantages to improve cyber risk
  management.
\newblock {\em The Geneva Papers on Risk and Insurance-Issues and
  Practice\/}~{\em 43}, 224--238.

\bibitem[\protect\citeauthoryear{Shevchenko, Jang, Malavasi, Peters, Sofronov,
  and Tr{\"u}ck}{Shevchenko et~al.}{2023}]{shevchenko2023}
Shevchenko, P.~V., J.~Jang, M.~Malavasi, G.~W. Peters, G.~Sofronov, and
  S.~Tr{\"u}ck (2023).
\newblock The nature of losses from cyber-related events: risk categories and
  business sectors.
\newblock {\em Journal of Cybersecurity\/}~{\em 9\/}(1), tyac016.

\bibitem[\protect\citeauthoryear{Stasinopoulos and Rigby}{Stasinopoulos and
  Rigby}{2008}]{stasinopoulos2008}
Stasinopoulos, D. and R.~Rigby (2008).
\newblock Generalized additive models for location scale and shape ({GAMLSS})
  in {R}.
\newblock pp.\  1–46.

\bibitem[\protect\citeauthoryear{Stasinopoulos, Rigby, Heller, Voudouris, and
  De~Bastiani}{Stasinopoulos et~al.}{2017}]{stasinopoulos2017}
Stasinopoulos, M.~D., R.~A. Rigby, G.~Z. Heller, V.~Voudouris, and
  F.~De~Bastiani (2017).
\newblock {\em Flexible regression and smoothing: using {GAMLSS} in {R}}.
\newblock CRC Press.

\bibitem[\protect\citeauthoryear{Sz{\'e}kely and Rizzo}{Sz{\'e}kely and
  Rizzo}{2013}]{szekely2013}
Sz{\'e}kely, G.~J. and M.~L. Rizzo (2013).
\newblock Energy statistics: A class of statistics based on distances.
\newblock {\em Journal of Statistical Planning and Inference\/}~{\em 143\/}(8),
  1249--1272.

\bibitem[\protect\citeauthoryear{Villase{\~n}or-Alva and
  Gonz{\'a}lez-Estrada}{Villase{\~n}or-Alva and
  Gonz{\'a}lez-Estrada}{2009}]{villasenor2009}
Villase{\~n}or-Alva, J.~A. and E.~Gonz{\'a}lez-Estrada (2009).
\newblock A bootstrap goodness of fit test for the generalized {Pareto}
  distribution.
\newblock {\em Computational Statistics \& Data Analysis\/}~{\em 53\/}(11),
  3835--3841.

\end{thebibliography}
\end{document}